\documentclass[11pt,a4paper]{article}
\pdfoutput=1

\usepackage{jheppub}
\usepackage{amsmath,amsfonts,mathrsfs,graphicx,bbm}

\def\S{{\mathbb S}}

\def\V{{\mathscr V}}
\def\SH{Q}
\def\SSH{{\mathcal Q}}
\def\W{{\mathscr W}}

\newcommand{\bea}{\begin{eqnarray}}
\newcommand{\eea}{\end{eqnarray}}

\newcommand{\alg}[1]{\mathfrak{#1}}

\newcommand{\psu}{\alg{psu}}

\newcommand{\el}{\nonumber\\}

\def\ads{{\rm AdS}_5\times {\rm S}^5}

\def\DH{{\Delta H}}

\newcommand{\KK}[1]{K^{\mathrm{#1}}}

\def\afQ{\widehat{\cal{Q}}}

\def\AdS{{\rm AdS}_5\times {\rm S}^5}

\def\am{{\rm am}}
\def\am0{{\rm am}_0}

\def\V{{\mbox v}}
\def\W{{ \mbox w}}
\def\U{{\mbox u}}

\title{The Quantum Deformed Mirror TBA II}

\author[a,1]{Gleb Arutyunov,}
\note{Correspondent fellow at Steklov
Mathematical Institute, Moscow.}
\author[b]{Marius de Leeuw,}
\author[a]{and Stijn J. van Tongeren}
\affiliation[a]{Institute for Theoretical
Physics and Spinoza Institute, Utrecht University, \\ Leuvenlaan \ 4, 3584 CE
Utrecht, The Netherlands}
\affiliation[b]{ETH Z\"urich, Institut f\"ur Theoretische Physik, \\Wolfgang-Pauli-Str.\ 27, CH-8093 Zurich, Switzerland}

\emailAdd{g.e.arutyunov@uu.nl}
\emailAdd{deleeuwm@phys.ethz.ch}
\emailAdd{s.j.vantongeren@uu.nl}

\abstract{We discuss the description of generic excited states in the quantum deformed $\AdS$ mirror thermodynamic Bethe ansatz and derive the associated Y-system. This Y-system shows an interesting new feature; it depends explicitly on the excited state under consideration. Similarly, it also depends on twisted boundary conditions. We construct the asymptotic solution of these TBA and Y-system equations by deriving the twisted transfer matrix for the quantum deformed Hubbard model and finding the deformed mirror bound state dressing phase. This asymptotic construction is insensitive to the precise nature of the deformation, and thereby provides a nontrivial check of the interesting new features which arise precisely at roots of unity.}

\begin{document}

\begin{flushright}\small{ITP-UU-12/34\\SPIN-12/32}\end{flushright}

\maketitle

\section{Introduction}

In this paper we continue our investigation of the thermodynamic Bethe ansatz (TBA) equations based on the quantum deformed $\AdS$ superstring S-matrix started in \cite{Arutyunov:2012zt}. This story builds on the remarkably successful application of integrability to the AdS/CFT correspondence \cite{Maldacena}, relating the free $\AdS$ superstring to maximally supersymmetric Yang-Mills (SYM) theory in the planar limit \cite{Arutyunov:2009ga,Beisert:2010jr}. In this setting, the spectrum of scaling dimensions of gauge invariant operators in SYM is equal to the spectrum of the $\AdS$ superstring - an integrable quantum field theory in finite volume.

The finite size ground state energy of the $\AdS$ superstring can be determined from the thermodynamics of its mirror model \cite{Zamolodchikov90,AF07}, resulting in a set of ground state TBA equations \cite{AF09b,BFT,GKKV09} based on the string hypothesis of \cite{AF09a}.\footnote{For recent reviews in the present context see e.g. \cite{Kuniba2,Bajnok:2010ke}.} These equations imply the corresponding Y-system \cite{GKV09} combined with intricate analytic properties studied in
\cite{Cavaglia:2010nm,Balog:2011nm}. Excited states can be described through a type of analytic continuation of the ground state TBA equations \cite{KP,DT96,BLZe,Fioravanti:1996rz,Arutyunov:2011uz}, explicitly done for various states in \cite{GKKV09,AFS09,BH10b,Sfondrini:2011rr,Arutyunov:2011mk, Arutyunov:2012tx}. The testing ground for these ideas has been the Konishi state, where the TBA approach \cite{AFS10,BH10a} agrees with L\"uscher's perturbative treatment \cite{BJ08,BJ09,LRV09,Janik:2010kd} and with the dual field theory up to five loops \cite{Sieg,Velizhanin:2008jd,Eden:2012fe}. The TBA and L\"uscher based results were recently extended to six and seven loop order respectively \cite{Leurent:2012ab,Bajnok:2012bz}, awaiting gauge theory verification.

The TBA approach has also been successfully applied to the cusp anomalous dimension in the context of Wilson loops \cite{Correa:2012hh,Drukker:2012de}, and there have been interesting developments in describing the TBA through a finite set of non-linear integral equations \cite{Gromov:2011cx,Balog:2012zt}\footnote{In fact, the approach of \cite{Gromov:2011cx} was used to compute the six-loop Konishi result of \cite{Leurent:2012ab}.}, see also \cite{Suzuki:2011dj}. In a specific limit, these last two topics merged nicely in \cite{Gromov:2012eu}. More generic boundary conditions also appear to be within reach of the $\AdS$ TBA \cite{Bajnok:2012xc}.

\begin{figure}[h]
\begin{center}
\includegraphics[width=0.95\textwidth]{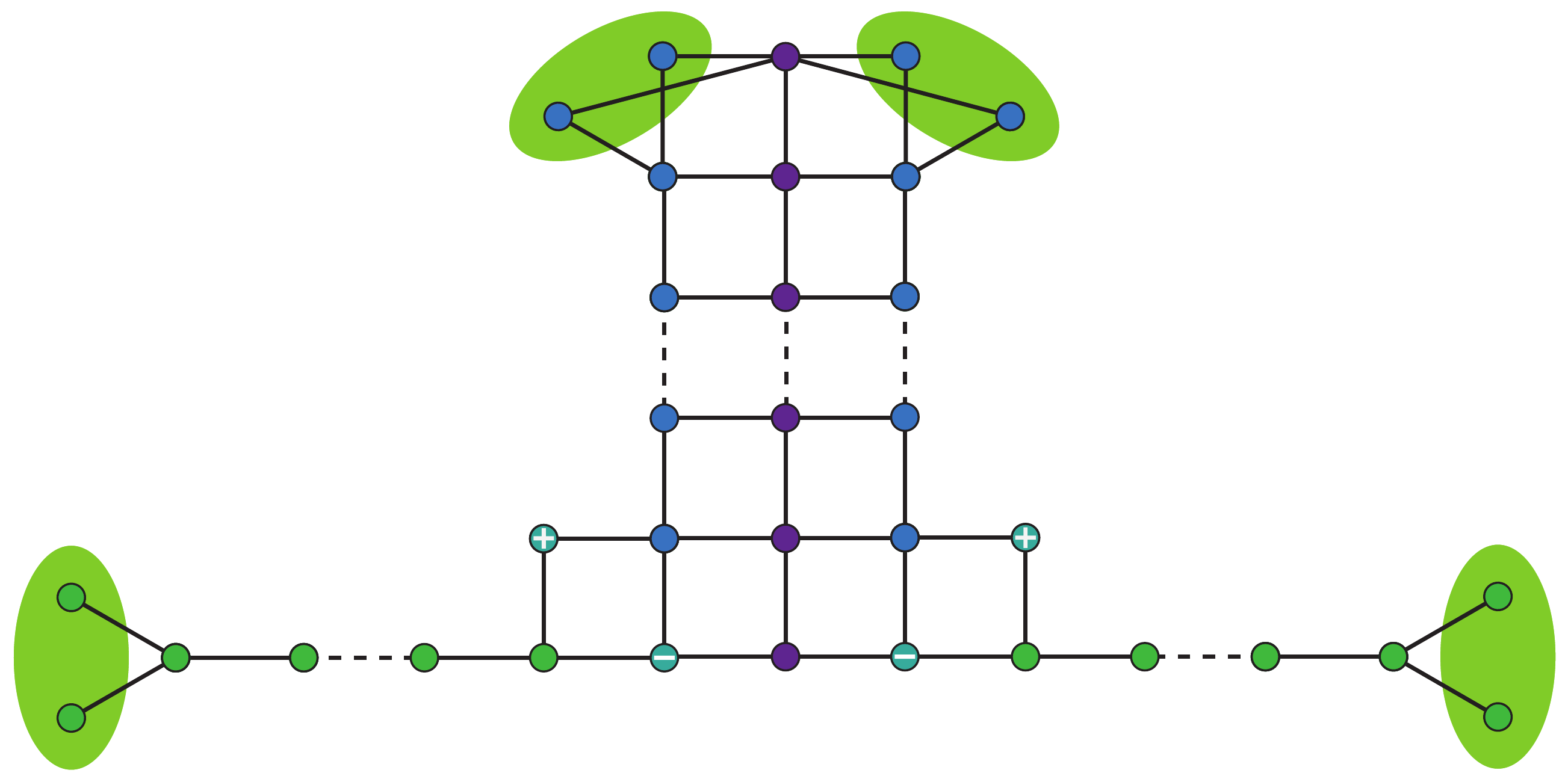}
\caption{The structure of the deformed TBA and Y-system. The green, teal (with $\pm$) and blue dots indicate what we call $Y_{M|w}$, $Y_{\pm}$, and $Y_{M|vw}$ functions respectively; their coupling is nearest neighbour apart from the indicated coupling near the corner. The two wings of this diagram represent the two copies of the deformed Hubbard model that are coupled via the momentum carrying $Y_Q$-functions (purple) of the mirror theory. Note that $Y_1$ couples to $Y_-$, but no $Y_Q$ couples to $Y_+$ in a local fashion. The lime-green bubbles signify the special state and twist dependent relation between $Y_{0|(v)w}$ and $Y_{k-1|(v)w}$ entering the Y-system.}
\label{fig:DefYsys}
\end{center}
\end{figure}

In our previous work, we derived ground state TBA equations based on the quantum deformed $\psu(2|2)^{\oplus 2}$ invariant S-matrix \cite{Beisert:2008tw,Hoare:2011wr} when the deformation parameter $q$ is a root of unity, more precisely $q = e^{ i \pi /k}$ with $k$ an integer greater than two. This S-matrix is conjectured  \cite{Hoare:2011fj,Hoare:2011wr,Hoare:2011nd} to interpolate between the S-matrix of the light-cone gauge fixed $\AdS$ superstring at $q =1$ and the S-matrix of the Pohlmeyer reduced $\AdS$ superstring \cite{Grigoriev:2007bu}\footnote{Recently the Pohlmeyer reduction procedure was also worked out for a string moving only in ${\rm AdS}_5$, which can be a more natural `vacuum' depending on the physical context \cite{Hoare:2012nx}.} in a relativistic limit where the coupling $g$ is taken to infinity. Because of the special nature of the deformation the TBA equations for this interpolating theory only contain a finite number of Y-functions with an interesting structure between them, as summarized schematically in figure \ref{fig:DefYsys}. We will discuss the actual equations in some detail in later sections.

We would like to extend these TBA equations to excited states in similar spirit to what we just described for the undeformed model. In the present paper we provide the necessary means to do so in the form of an asymptotic solution. This asymptotic solution is a solution of the excited state TBA equations in the limit where the momentum carrying $Y_Q$ functions are small. Once found by for example the contour deformation trick \cite{Arutyunov:2011uz}, these excited state TBA equations hold also outside this limit. Apart from the finite number of Y-functions and deformed integration kernels entering in the equations, this story mirrors the undeformed one perfectly up to this point. However, the fact that we have only finitely many Y-functions - a `boundary' - introduces an important difference.

In the undeformed case, we can derive a set of so-called Y-system relations between the Y-functions which appear to be of the same form for any excited state; they are universal \cite{Zamolodchikov:1991et}, see \cite{GKV09} in the present context. Similarly, the undeformed Y-system is universal for all twisted boundary conditions in the string theory \cite{arXiv:1009.4118,Gromov:2010dy,deLeeuw:2011rw,arXiv:1108.4914,dLvT2}. For our deformed model on the other hand, the Y-system is no longer universal in either sense. Near the boundaries of figure \ref{fig:DefYsys} the Y-system appears to pick up a mild dependence on (the excitation numbers of) the state under consideration as well as the twisted boundary conditions. For twisted boundary conditions this can already be observed for the XXZ spin chain \cite{Takahashi:1972}, however we have not encountered this state dependence before. As this effect appears to depend crucially on the combination of the deformation with the fact that we have a nested system, perhaps this is not too surprising.

As we will discuss in detail right after the introduction, this dependence comes about in two ways for either cause. Firstly the inverse relation between $Y_{0|(v)w}$ and $Y_{k-1|(v)w}$ is not preserved by chemical potentials associated to twisted boundary conditions. This is because the $0|(v)w$ and $k-1|(v)w$ strings have wildly different charges rather than the opposite ones which would be required to preserve the inverse relation. Secondly, while twisted boundary conditions result in chemical potentials that are annihilated by the discrete Laplace operator used to derive the Y-system, near the boundary we necessarily apply a slightly different operator which leaves a chemical potential dependence. Next, the state dependence comes in firstly because the driving terms for $Y_{0|(v)w}$ and $Y_{k-1|(v)w}$ naturally do not preserve their inverse relation; while resulting in kernels with opposite signs, their S-matrices are in fact anti-inverse rather than inverse. Furthermore, while the kernels are annihilated by the boundary operator we apply to obtain the Y-system, this only implies that the driving terms should be annihilated up to a constant term. This constant term turns out to be another minus sign in the game. These special relations between $Y_{0|(v)w}$ and $Y_{k-1|(v)w}$ are emphasized by the lime-green bubbles in figure \ref{fig:DefYsys}. This interesting dependence is of course reproduced by our asymptotic solution.

The main results of this paper are the explicit derivation of the excited state Y-system and the \emph{independent} derivation of its asymptotic solution, which showcases the highly involved structure of this interesting model while at the same time providing a nontrivial check of our string hypothesis. In addition to this we discuss the relativistic limit of our TBA equations, where they reduce to the TBA equations based on the conjectured S-matrix for the Pohlmeyer reduced theory. Of course in this limit the TBA equations simplify considerably. Most notably, the Y-functions become meromorphic functions on the $u$-plane and all nontrivial kernels including the dressing phase disappear completely from the simplified TBA equations.\footnote{The information contained in the kernels is of course still contained implicitly in the asymptotics of the Y-functions.}

We have also carefully derived an explicit expression for the $q$-deformed mirror bound state dressing phase, and proved that it satisfies properties completely analogous to the undeformed case. Along the way we also show that the $q$-deformed transfer matrix used to construct the asymptotic solution can be obtained via fusion in the generating functional approach, and discuss the twisted ground state solution of the deformed model which takes a form very similar to the undeformed theory.

This paper is organized as follows. In the next section we discuss the dependence of the Y-system on twists and excitation numbers, give a practical set of simplified TBA equations and the Y-system where these features have been explicitly taken into account, and discuss the mapping of the Y-system to the T-system in detail. Then in section \ref{sec:Transfermatrix} we derive the transfer matrix for the $q$-deformed theory with twists, which is used in section \ref{sec:Asymptoticsolution} to construct the asymptotic solution of the TBA equations and Y-system. We then discuss the relativistic limit of our TBA equations in section \ref{sec:Relativisticlimit}, and finish up with a discussion of our results and outlook in section \ref{sec:conclusion}. Various background material and technical details are presented in appendices; in particular we give a complete derivation of the deformed mirror bound state dressing phase in appendix \ref{app:dressing phase}.

\section{The Y-system for excited states and twists}
\label{sec:Ysysgeneral}

Typically the Y-system associated to a set of TBA equations is universal in the sense that it is the same for all excited states and independent of chemical potentials associated to symmetries of the theory. This is not true for the present theory, a fact which is intimately related to having a finite number of TBA equations.\footnote{While atypical, chemical potentials (magnetic fields) already appear in the Y-system of the XXZ spin chain at roots of unity \cite{Takahashi:1972}.} Let us begin by showing the effect of considering an excited state at the level of the canonical TBA equations, on the simplified TBA equations and associated Y-system. We will work this out explicitly for the simple case of $w$-strings.

\subsection{Excited states in the deformed mirror TBA}

The canonical ground state TBA equations for $w$ strings are given by \cite{Arutyunov:2012zt}
\begin{equation}
\label{eq:cTBAw0}
\log{Y_{M|w}^{(r)}}= \,\log\left(1+\tfrac{1}{Y_{N|w}^{(r)}}\right)\star K_{NM} +\log\frac{1-\frac{1}{Y_-^{(r)}}}{1-\frac{1}{Y_+^{(r)}}}\,\hat{\star}\,K_M\, ,
\end{equation}
where $r=\pm$ is a label referring to the two copies of $\mathfrak{su}(2|2)$, and we have not eliminated the Y-function corresponding to negative parity strings yet. Hence, it is important to note that the indices $M$ and $N$ are general and run over the length one to $k-1$ positive parity strings, and the length one negative parity string. Denoting the negative parity length one string as type $0$ we note that the $S$-matrices the above kernels are derived from have the properties\footnote{The explicit definition and properties of the kernels and S-matrices can be found in \cite{Arutyunov:2012zt}.}
\begin{equation}
\label{eq:basicSmatrels}
S^{k-1} S^{0} = -1 \, , \, \, \, \, S^{Mk-1}S^{M0} = 1.
\end{equation}
This immediately implies $Y_{0|w}^{(r)} = \left(Y_{k-1|w}^{(r)}\right)^{-1}$ for the ground state. For excited states however, we should be more careful.

In general, the TBA equations for an excited state are the same as those for the ground state, up to a set of driving terms \cite{DT96,BLZe}. These driving terms depend on the state under consideration and are different for each particle type's TBA equation. Let us for simplicity assume that we have a state with $K_w$ roots of $1-Y_-$ that enter the TBA equations and no further roots. Considering the above form of the TBA equations, this would result in driving term contributions of the form
\begin{equation}
\sum_{i=1}^{K_w} \log S^M(r_i^{(r)} - u - \tfrac{i}{g})\, ,
\end{equation}
in the equation for $\log Y^{(r)}_{M|w}(u)$. Because of the first relation in eq. \eqref{eq:basicSmatrels}, this results in a relative sign between the driving terms for $\log Y_{k-1|w}$ and $\log Y_{0|w}$, exactly in line with their Y-functions being inverse, as for the ground state. However, due to the minus sign in the first of eqs. \eqref{eq:basicSmatrels} we also get an extra factor of $K_w i \pi$.\footnote{Of course the reader can argue that an S-matrix-like driving term $\log S(u,v)$ based on a kernel $K$ is defined up functions of $v$ only. This ambiguity should then be fixed by comparison to an asymptotic solution for the Y-functions. We do not have a direct asymptotic form for $Y_{0|w}$, making this impossible. However, let us firstly note that the chosen driving terms are the completely natural choice. Moreover, as we will discuss below the resulting equations are compatible with the available asymptotic form for the other Y-functions, while other potential forms of the driving terms are not. Hence there is no true ambiguity.} This means that while
\begin{equation}
Y_{0|w}^{(r)} Y_{k-1|w}^{(r)} =1\, \, \, \, \, \, \mbox{for even}\, \, K_w\, ,
\end{equation}
we have
\begin{equation}
Y_{0|w}^{(r)} Y_{k-1|w}^{(r)} = -1 \, \, \, \, \, \, \mbox{for odd}\, \, K_w\, .
\end{equation}
Now let us come back to our original assumption. If we were to consider additional roots of $Y_\pm$ or $1-Y_+$, clearly $K_w$ above should be replaced by the total number of all these roots, taking relative signs between driving terms into account. Furthermore, due to the plus sign in the second relation in eq.\eqref{eq:basicSmatrels} additional roots of $1+Y_{M|w}$ do not further affect this relation.

Next, these driving terms also play a direct and important role in the simplified TBA equations and Y-system for $Y_{k-1|w}$. Conventionally, the TBA equations can be rewritten in simplified form by noting identities between various kernels. As convolutions with convoluted kernels turn into convolutions with the standard kernel $s$, the driving terms in the simplified TBA equations turn into $\log S$ terms, where $S$ is the standard $S$-matrix associated to $s$
\begin{equation}
s = \frac{1}{2 \pi i} \frac{d}{du} \log S \, .
\end{equation}
Not surprisingly, when acting with $s^{-1}$ to obtain the Y-system these driving terms are annihilated.\footnote{$s^{-1}$ is defined below in eqn. \eqref{eq:sinv}.} Up to potential subtleties with branch cuts, these identities between nontrivial S-matrices and the subsequent annihilation of the remainder by $s^{-1}$ can be summarized by the statement that these S-matrices satisfy the discrete Laplace equation
\begin{equation}
\frac{S^M (u,v+i/g)S^M (u,v-i/g)}{S^{M+1}(u,v)S^{M-1} (u,v)} =1\, ,
\end{equation}
which holds for any of the relevant S-matrices in what follows. Since we have a boundary however, the identities we use there are equivalent to getting a final contribution of
\begin{equation}
\label{eq:bndrydrivingterm}
\tfrac{1}{2} \log \frac{S^{k-1}(u,v+i/g)S^{k-1} (u,v-i/g)}{S^{k-2} (u,v)} = \tfrac{1}{2} \log S^{k} (u,v)
\end{equation}
in the simplified TBA equation for $Y_{k-1|w}$. Of course $S^{k}(u,v)$ is typically constant such that its associated kernel vanishes. However, with the auxiliary problem being $2k$ periodic, $S^k$ is not quite one, but rather minus one. This means that the simplified TBA equation for $Y_{k-1|w}$ gets an extra factor of $K_w i \pi/2$. Repeating the above analysis for the canonical TBA equations for $vw$ strings,
\begin{equation}
\label{eq:cTBAvw0}
\log{Y_{M|vw}^{(r)}}
=\log\left(1+\tfrac{1}{Y_{N|vw}^{(r)}}\right)\star K_{NM} + \log\frac{1-\frac{1}{Y_-^{(r)}}}{1-\frac{1}{Y_+^{(r)}}}\,\hat{\star}\,K_M - \log\left(1+Y_Q\right)\star K^{QM}_{xv}\,,
\end{equation}
it is easy to see that the relation between $Y_{0|vw}$ and $Y_{k-1|vw}$ should be sensitive to both the roots associated to $Y_\pm$ and those to associated to $Y_Q$, as $S^{Q0}_{xv}S^{Qk-1}_{xv}=-1$ as well. Let us denote the number additional roots due to $Y_Q$ by $K_{vw}$.

At this point we might wonder whether the number of these roots effectively follows a pattern. To understand and motivate this, we should realize that physically speaking the difference between the driving terms for the $0|(v)w$ and $k-1|(v)w$ string comes about because of their different charges; they feel the presence of the excited state differently. This hints that the number of roots could be directly related to the charge of the state, in particular to their excitation numbers.

To continue along this track, let us start by considering generic states in the extensively studied $\mathfrak{sl}(2)$ sector of the undeformed theory. Based on various studies of the TBA equations for excited states in this theory \cite{GKKV09,AFS09,BH10b,AFunpublished}, it seems reasonable to assume that here the number of relevant roots is restricted to those of $1+Y_Q$ corresponding to the fundamental excitations in the string theory. This would mean that $K_{vw} = K^{\rm I}$ and $K_w = 0$ in this sector. Next, if we consider states with nonzero $K^{\rm II}_{(r)}$ with real momenta such as those studied in \cite{Sfondrini:2011rr} the natural generalization of this statement is that $K_{vw} = K^{\rm I}$ and $K_w = K^{\rm II}_{(r)}$.\footnote{Our excitation numbers refer to the $\mathfrak{sl}(2)$ grading.} At the level of the number of roots this simple picture is not quite true anymore for for example an orbifolded magnon \cite{arXiv:1009.4118} or states with complex momenta in the $\mathfrak{su}(2)$ sector \cite{Arutyunov:2011mk}. However, we should note that the relation between the Y-functions only depends on whether $K_w$ and $K_w + K_{vw}$ are even or odd, and that in this sense any currently investigated state matches the described pattern. Assuming this picture not to change qualitatively in our deformed model, we then have the natural conjecture that in general
\begin{equation}
\label{eq:excYreln}
\begin{aligned}
Y_{0|w} Y_{k-1|w} & = (-1)^{K^{\rm II}_{(r)}}\, , \\
Y_{0|vw} Y_{k-1|vw} & = (-1)^{K^{\rm I}+K^{\rm II}_{(r)}}\, .
\end{aligned}
\end{equation}

We might wonder whether or not the excitation number $K^{\rm III}_{(r)}$ should make an appearance. To understand let us first look at the simple case of $w$ strings again. In essence there the above story tells us that the `length of the spin chain' associated to a particular particle type (nested level) determines the appearance of the minus sign. However, then we would immediately expect a minus sign for $vw$ strings depending on whether $K^{\rm I}$ is even or odd, apparently contradicting the above. The reason for the appearance of $K^{\rm II}_{(r)}$ in this picture is that in deriving the 'canonical' TBA equations for $vw$ strings the canonical TBA equations for $w$ strings were used to exchange the presence of infinitely many $Y_{M|w}$ functions for the presence of $Y_\pm$ \cite{AF09a,Arutyunov:2012zt}. If this had not been done, the direct dependence would be on $K^{\rm I}$ only, where the dependence on $K^{\rm II}_{(r)}$ would come in when eliminating the $Y_{M|w}$ functions from the simplified TBA or Y-system. From this point of view the presence of $w$ roots (nonzero $K^{\rm III}_{(r)}$) cannot result in any potential minus signs because there is no deeper nested level.

While the precise form of the general relations \eqref{eq:excYreln} is a conjecture, it is completely natural to expect simple relations of this type by the physical picture that they are caused by the difference in charges of the respective string solutions. Of course, this conjecture is also supported by the asymptotic Y-system we will discuss in section \ref{sec:Asymptoticsolution}. Finally, state dependence is clearly present by the above discussion, regardless of the precise form.

\subsection{Boundary effects for boundary conditions in the mirror trick}

As it turns out, the Y-system for our mirror model also depends explicitly on boundary conditions in the original model, which manifest themselves as chemical potentials in the TBA equations \cite{Zamolodchikov90,Bajnok:2004jd}. Normally such chemical potentials drop out of the Y-system; here their effect does not completely cancel near the boundary of the Y-system.

\begin{figure}[h]
\begin{center}
\includegraphics[width=\textwidth]{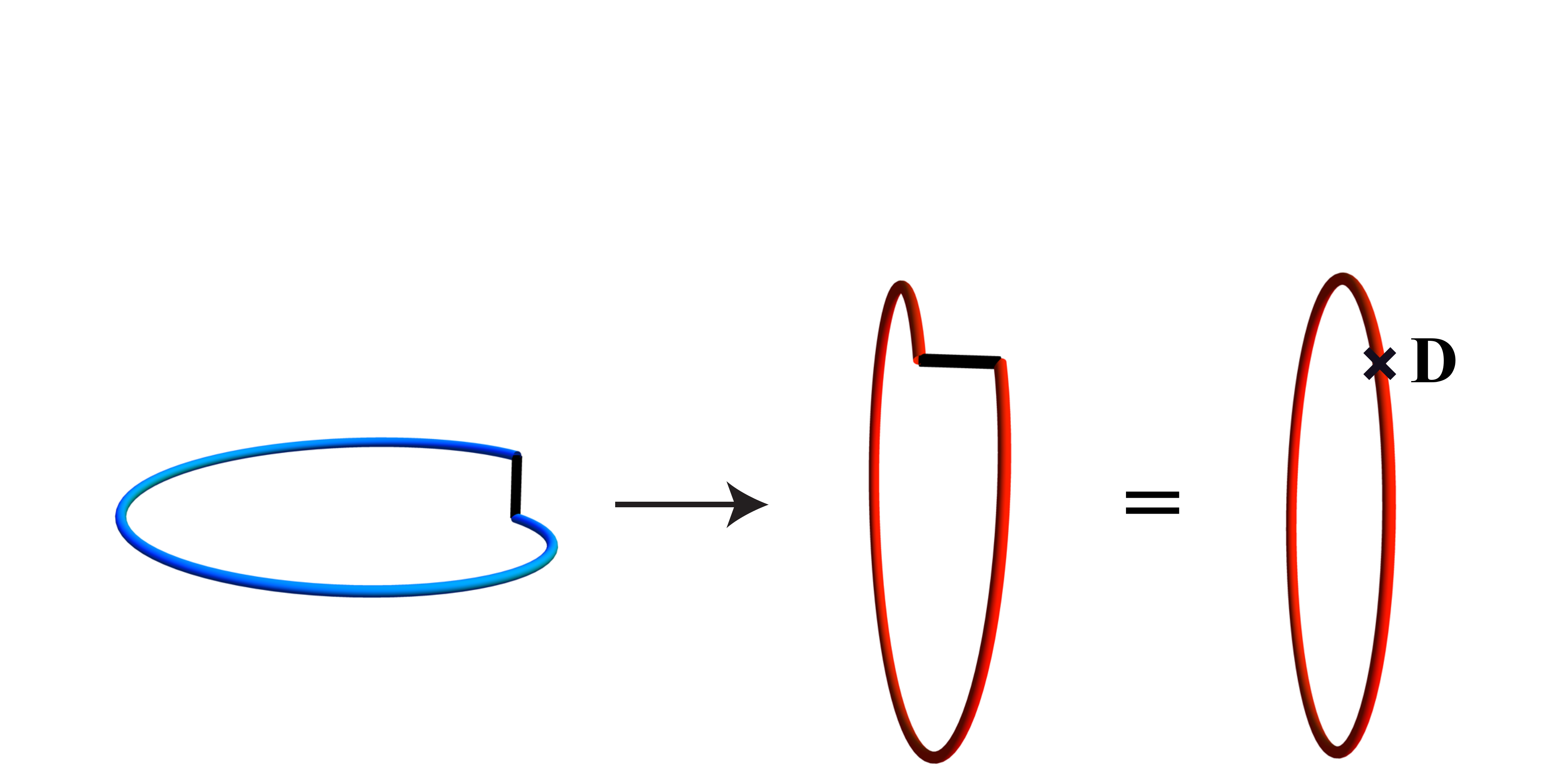}
\caption{Quasi-periodic boundary conditions give a defect operator in the mirror theory. The quasi-periodic boundary condition denoted by the black discontinuity on the circle of the original theory (blue), turns into a discontinuity in the mirror time evolution (red), which is equivalent to insertion of a defect operator $D$ in the partition function.}
\label{fig:bcmirrortrick}
\end{center}
\end{figure}

Incorporating nontrivial boundary conditions in the original model corresponds to insertion of a defect operator in the path integral for the mirror model, as illustrated in figure \ref{fig:bcmirrortrick}.
This operator results in the appearance of chemical potentials on the right hand side of the canonical TBA equations, worked out carefully in the present context in \cite{arXiv:1108.4914,dLvT2}, see also \cite{deLeeuw:2011rw}. However, as the negative parity string has charge one in appropriate units compared to the charge $k-1$ of the length $k-1$ string, their chemical potentials are not quite opposite but rather sum to $k$ units of chemical potential. This means the presence of a chemical potential results in a further modified relation between $Y_{0|(v)w}$ and $Y_{k-1|(v)w}$, namely
\begin{equation}
Y_{0|(v)w}Y_{k-1|(v)w} = (-1)^{K^{\rm{II}}_{(\alpha)}(+K^{\rm{I}})} e^{- k \mu} \, .
\end{equation}
Moreover, when deriving the simplified TBA equation for $Y_{k-1|(v)w}$ we are left with a chemical potential contribution just as we were left with an S-matrix as in eq. \eqref{eq:bndrydrivingterm}. This results in a total factor of $-\tfrac{1}{2}(k \hspace{0.5pt} \mu + i \pi K^{\rm{II}}_{(\alpha)}(+K^{\rm{I}}))$ in the simplified TBA equation for $Y_{k-1|(v)w}$.

Intuitively these boundary chemical potentials as well as the minus signs discussed above come about because while the $0|(v)w$ and $k-1|(v)w$ strings scatter inversely in the thermodynamic limit (as if they were each others anti-particles), they do not have opposite charges.

\subsection{The simplified TBA equations and Y-system}

The above discussion makes it clear that the simplified TBA equations and Y-system pick up an explicit dependence on the chemical potential and (the excitation numbers of) the state under consideration, in addition to conventional driving terms that generically still enter the simplified TBA equations for excited states. Part of this dependence comes about through the boundary associated with having finitely many TBA equations, while the other part comes in explicitly by eliminating $Y_{0|(v)w}$ from the TBA equations. We could choose not to eliminate $Y_{0|(v)w}$ from the TBA equations, but since this would not remove the dependence on excitation numbers and chemical potential completely in any case, and we would have to keep extra careful track of the origin of driving terms, it seems best to eliminate $Y_{0|(v)w}$. This way the TBA equations can be treated as in the undeformed case, with an asymptotic solution for the remaining Y-functions and the contour deformation trick applying directly and straightforwardly. The only modification is that we additionally have the manifest dependence on the excitation numbers and chemical potential. Let us note again that this dependence does derive properly from the contour deformation trick and the full set of canonical TBA equations as discussed above; it is simply easier to manifest it before continuing.

With this in mind we can write down the simplified TBA equations for excited states with nonzero chemical potentials, up to standard $\log S$ terms following from the analytic properties of the \emph{independent} Y-functions.\footnote{For $Y_\pm$ we still write the canonical TBA equations, in principle they can be brought to so-called hybrid form \cite{AFS09} but we will not pursue this here.} The full mirror model has two sets of auxiliary equations labeled by an index
$r=\pm$. Suppressing the index, for one copy these are given by
\begin{align}
\log{Y_{M|vw}} = \, & \log{(1+Y_{M+1|vw})(1+Y_{M-1|vw})} \star s -
\log\left(1+Y_{M+1}\right)\star s + \delta_{M,1}
\log{\frac{1-Y_-}{1-Y_+}}  \hat{\star}  s \nonumber\\
\log{Y_{k-2|vw}}  = \, & \log{(1+Y_{k-3|vw})(1+Y_{k-1|vw})(1+e^{\chi_{vw}}Y_{k-1|vw})} \star s  - \log\left(1+Y_{k-1}\right)\star s\, ,\nonumber \\
\log{Y_{k-1|vw}}  = \, & \log{(1+Y_{k-2|vw})} \star s - \log\left(1+Y_{k}\right)\star s - \tfrac{\chi_{vw}}{2}\, ,\nonumber \\
\log{Y_{M|w}}  = \, &\log{(1+Y_{M+1|w})(1+Y_{M-1|w})} \star s + \delta_{M,1}
\log{\frac{1-Y^{-1}_-}{1-Y^{-1}_+}}  \hat{\star}  s\, ,\\
\log{Y_{k-2|w}}  = \, &\log{(1+Y_{k-3|w})(1+Y_{k-1|w})(1+e^{\chi_{w}}Y_{k-1|w})} \star s\, , \nonumber\\
\log{Y_{k-1|w}}  = \, &\log{(1+Y_{k-2|w})} \star s - \tfrac{\chi_{w}}{2}\, , \nonumber \\
\log{Y_{\pm}} =\, & - \mu_y -\log\left(1+Y_{Q}\right)\star K^{Qy}_{\pm} +
\log\frac{1+Y^{-1}_{M|vw}}{1+Y^{-1}_{M|w}}\star K_{M}+
\log\frac{1+Y_{k-1|vw}}{1+Y_{k-1|w}}\star K_{k-1} \, ,\nonumber
\end{align}
where
\begin{align*}
\chi^{(r)}_{vw} & = k \hspace{1pt} \mu^{(r)}_{1|vw} + i \pi (K^{\rm
I}+K^{\rm II}_{(r)})\, ,\\
\chi^{(r)}_{w} & = k \hspace{1pt} \mu^{(r)}_{1|w} + i \pi K^{\rm
II}_{(r)}\,.
\end{align*}
The chemical potentials we consider derive from twisted boundary conditions in the original model. As in the undeformed model \cite{dLvT2} they are naturally parametrized in terms of the twists $\alpha_{r}$ and $\beta_{r}$ as summarized in table \ref{tab:chempot}.
\begin{table}[h!]
\begin{center}
\begin{tabular}{|c|c|}
\hline  $A$ & $\mu_A^{(r)}$ \\
\hline
\hline $M|w$   & $2 i M \beta_r$  \\
\hline $M|vw$    & $2 i M \alpha_r$  \\
\hline $y$    & $i(\alpha_r - \beta_r)$ \\
\hline $Q$   & $i Q (\alpha_+ + \alpha_-)$   \\
\hline
\end{tabular}
\end{center}
\caption{Chemical potentials corresponding to twists for the particle content of the deformed mirror
TBA.}
\label{tab:chempot}
\end{table}

The momentum carrying $Y_Q$ functions couple two sets of the above equations through
\begin{align}
\hspace{-60pt}
\log Y_1= \,& \log\frac{Y_2}{1+Y_2} \star s +
\log\prod_{r=\pm}\left(1-\tfrac{1}{Y^{(r)}_-}\right)\star s -\check{\Delta}
\check{\star} s\, , \nonumber\\
\log Y_Q= \,&  \log\frac{{Y_{Q+1}Y_{Q-1}}}{(1+Y_{Q-1})(1+Y_{Q+1})}\star s + \log
\prod_{r=\pm} \left(1+\tfrac{1}{Y^{(r)}_{Q-1|vw}}\right) \star s\, , \\
\log Y_k = \,&\log\frac{Y_{k-1}^2}{1+Y_{k-1}}\star s+
\log\prod_{r=\pm}\left(1+\tfrac{1}{Y^{(r)}_{k-1|vw}}\right)\left(1+\tfrac{
e^{-\chi^{(r)}_{vw}}}{Y^{(r)}_{k-1|vw}}\right) \star s \, .\nonumber
\end{align}
Note that in the equation for $Y_k$ just above we might expect a
chemical potential term because $\mu_Q$ is not in the kernel of $\delta_{Qk} -
2 s \delta_{Qk-1}$. However, the resulting factor of $2\mu_{1}$ is exactly cancelled by the chemical potentials for a $(+)$ and a $(-)$ $1|vw$ string, cf. table \ref{tab:chempot}, whose canonical TBA equations are used to obtain the simplified equation for $Y_k$ as discussed in appendix E of \cite{Arutyunov:2012zt}. The Y-system can be immediately read off from these equations by applying $s^{-1}$ to them\footnote{For $Y_\pm$ the story is more subtle. $Y_-$ has a Y-system equation obtained via TBA identities, while $Y_+$ has no local Y-system equation.}, where
\begin{equation}
\label{eq:sinv}
f \circ s^{-1}(u) \equiv \lim_{\epsilon\to 0} [f(u+i/g-i \epsilon) + f(u-i/g+i \epsilon)] \, .
\end{equation}
Explicitly the Y-system is given by

\paragraph{$w$ strings}
\begin{align}
Y_{1|w}^+ Y_{1|w}^- & =(1+Y_{2|w})\left(\frac{1-Y_-^{-1}}{1-Y_+^{-1}}\right)^{\theta(u_B-|u|)}  \\
Y_{M|w}^+ Y_{M|w}^- & =(1+Y_{M-1|w})(1+Y_{M+1|w}) \, , \\
Y_{k-2|w}^+ Y_{k-2|w}^- & = (1+Y_{k-3|w})(1+Y_{k-1|w})(1 + e^{\chi_w}Y_{k-1|w}) \, ,\label{eqn;Zw1}\\
Y_{k-1|w}^+ Y_{k-1|w}^- & = e^{-\chi_w}(1+Y_{k-2|w}) \, .\label{eqn;Zw2}
\end{align}
\paragraph{$vw$ strings}
\begin{align}
Y_{1|vw}^+ Y_{1|vw}^- & =\frac{1+Y_{2|vw}}{1+Y_2}\left(\frac{1-Y_-}{1-Y_+}\right)^{\theta(u_B-|u|)}  \\
Y_{M|vw}^+ Y_{M|vw}^- & =\frac{(1+Y_{M-1|vw})(1+Y_{M+1|vw})}{1+Y_{M+1}}\, ,\\
Y_{k-2|vw}^+ Y_{k-2|vw}^- & = \frac{(1+Y_{k-3|vw})(1 + Y_{k-1|vw})
(1 +e^{\chi_{vw}} Y_{k-1|vw})}{1+Y_{k-1}}\, ,\\
Y_{k-1|vw}^+ Y_{k-1|vw}^- & = e^{-\chi_{vw}}\frac{1+Y_{k-2|vw}}{1+Y_k}\, .
\end{align}

\paragraph{$y$ strings}
\begin{equation}
Y_{-}^{+}Y_{-}^{-} = \frac{1+Y_{1|vw}}{1+Y_{1|w}}\frac{1}{1+Y_1}\, .
\end{equation}

\paragraph{Q particles\\}
\begin{align}
\frac{Y_1^+ Y_1^-}{Y_2} & = \frac{\displaystyle{\prod_{r=\pm}}
\left(1-\frac{1}{Y^{(r)}_{-}}\right)}{1+Y_2}\, , \, \, \, \, \, \mbox{for} \, \,
|u|<u_b\, .\\
\frac{Y_Q^+ Y_Q^-}{Y_{Q+1}Y_{Q-1}} & = \frac{\displaystyle{\prod_{r=\pm}} \Bigg(
1+\frac{1}{Y_{Q-1|vw}^{(r)}}\Bigg)}{(1+Y_{Q-1})(1+Y_{Q+1})} \, , \\
\frac{Y_k^+ Y_k^-}{Y_{k-1}^2} & = \frac{\displaystyle{\prod_{r=\pm}}
\Bigg(1 +\frac{1}{Y_{k-1|vw}^{(r)}}\Bigg)
\Bigg(1 +\frac{e^{-\chi_{vw}}}{Y_{k-1|vw}^{(r)}}\Bigg)}{1+Y_{k-1}}\,.
\end{align}

The above equations have an explicit dependence on $\chi$ that cannot be removed by simply redefining the Y-functions. However, let us note that by the redefinition
\begin{equation}
Y_{k-1|(v)w} \rightarrow e^{-\tfrac{\chi_{(v)w}}{2}} Y_{k-1|(v)w}\,
\end{equation}
the above equations do become manifestly symmetric in $\chi$, and in particular in the twist. Since we would expect the spectrum to be symmetric in the twist, this is only natural.

\subsection{T-system}

\begin{figure}[h]
\begin{center}
\includegraphics[width=\textwidth]{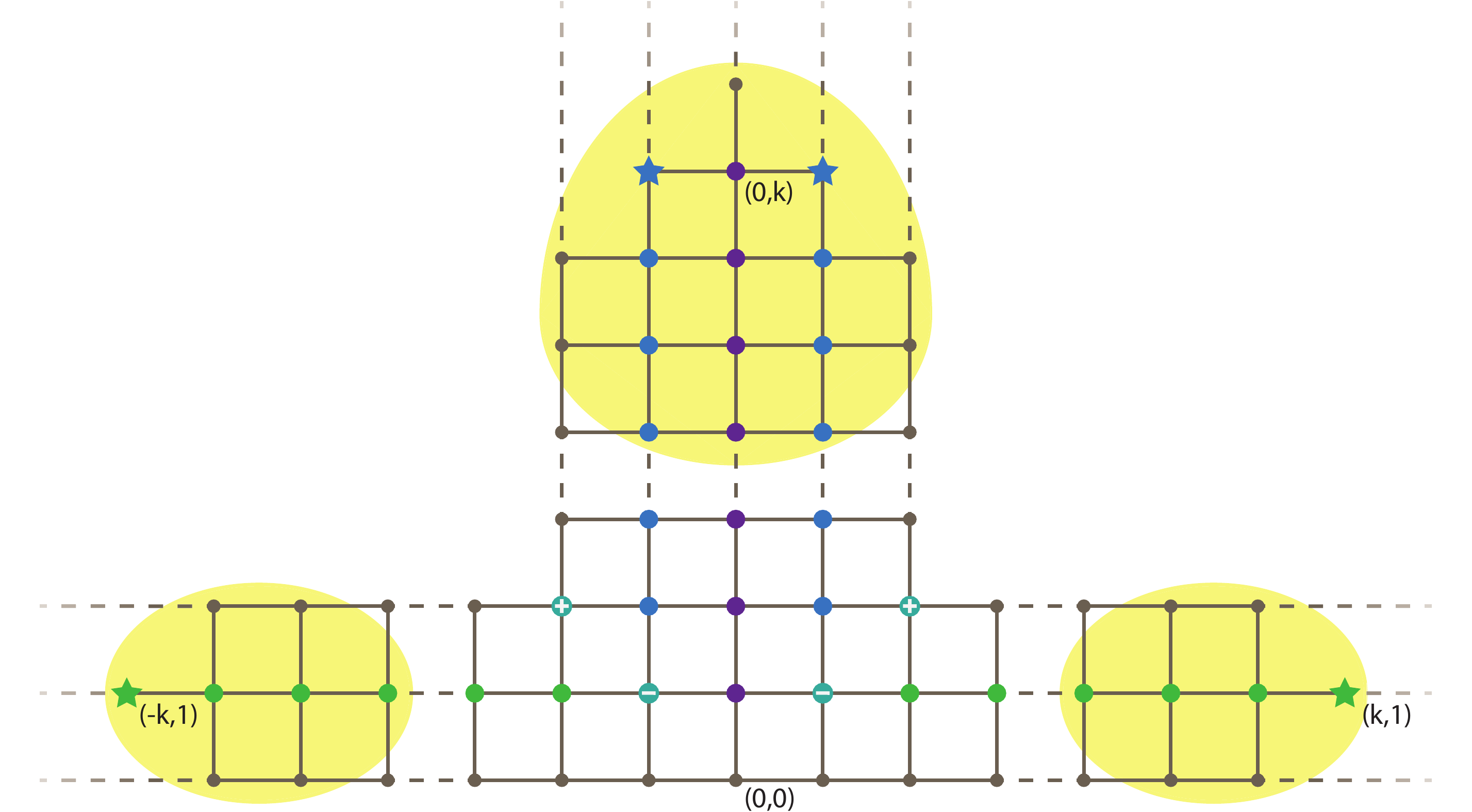}
\caption{The T-hook lattice for T-functions satisfying the Hirota
equations. There is a T-function $T_{a,s}$ indicated by a gray dot at any
lattice site $(s,a)$. The bigger coloured dots cover underlying gray dots and
signify the Y-functions constructed out of nearest neighbour T-functions; the
stars denote the special Y-functions which have no direct construction in terms
of standard T-functions. By eliminating the special Y-functions, the
consistency of the Y-system near the boundary is equivalent to a set of
non-standard additional relations between the T-functions in each respective
yellow bubble. These relations allow us to consider a truncation of the in
principle infinite lattice of T-functions satisfying the Hirota equations.}
\label{fig:Thook}
\end{center}
\end{figure}

It is useful to express the Y-functions in terms of so-called T-functions. In the undeformed case the Y-system equations then become equivalent to a set of Hirota equations for these T-functions. Here we will study what happens in the quantum deformed model.

As in the undeformed case we relate our Y-system to the T-system by making the following identifications \cite{KP,Kuniba:1993cn,Krichever:1996qd},\cite{GKV09,AFS09}
\begin{align}\label{eqn;Tsystem}
& Y^{(\pm)}_{M|w} = \frac{T_{1,\pm(M+2)}T_{1,\pm M}}{T_{0,\pm(M+1)}T_{2,\pm(M+1)}}, &&
Y^{(\pm)}_{M|vw} = \frac{T_{M+2,\pm1}T_{M,\pm1}}{T_{M+1,0}T_{M+1,\pm2}}, &&
Y_{M} = \frac{T_{M,1}T_{M,-1}}{T_{M+1,0}T_{M-1,0}}\\
& Y^{(\pm)}_{+} = -\frac{T_{2,\pm1}T_{2,\pm3}}{T_{1,\pm2}T_{1,\pm3}}, &&
 Y^{(\pm)}_{-} = - \frac{T_{0,\pm1}T_{2,\pm1}}{T_{1,0}T_{1,\pm2}}.
\end{align}
In the undeformed limit, the Y-system equations are equivalent to the Hirota equations
\begin{equation}
T_{a,s}^+T_{a,s}^- = T_{a+1,s}T_{a-1,s} + T_{a,s+1}T_{a,s-1}.
\end{equation}
plus the boundary conditions that $T_{a,s}$ vanishes if both $a$ and $|s|$ are greater than two. These boundary conditions mean that the T-functions lie on a so-called fat hook, as illustrated in figure \ref{fig:Thook}. Away from $M=k-1,k-2$ this identification and equivalence also immediately work for our $q$-deformed model. However, at the boundary where $a,s=k,k+1$ we obviously find special relations. In principle we can eliminate the T-functions that do not satisfy the standard Hirota equations to obtain a set of standard T-functions that solve the Hirota equations\footnote{Since we have only a finite number of T-functions by construction, here we mean that they solve the Hirota equations in the `forward' sense, going outward from the origin of the lattice.}, plus a set of complicated additional relations signifying the closure of Y-system at $M=k-1,k-2$. This point of view is illustrated in figure \ref{fig:Thook}, where these extra relations are schematically indicated by the yellow shading, simultaneously marking the boundary of our T-system. Note that these relations are not between nearest neighbours only, as the Hirota equations are.

\begin{figure}[h]
\begin{center}
\includegraphics[width=\textwidth]{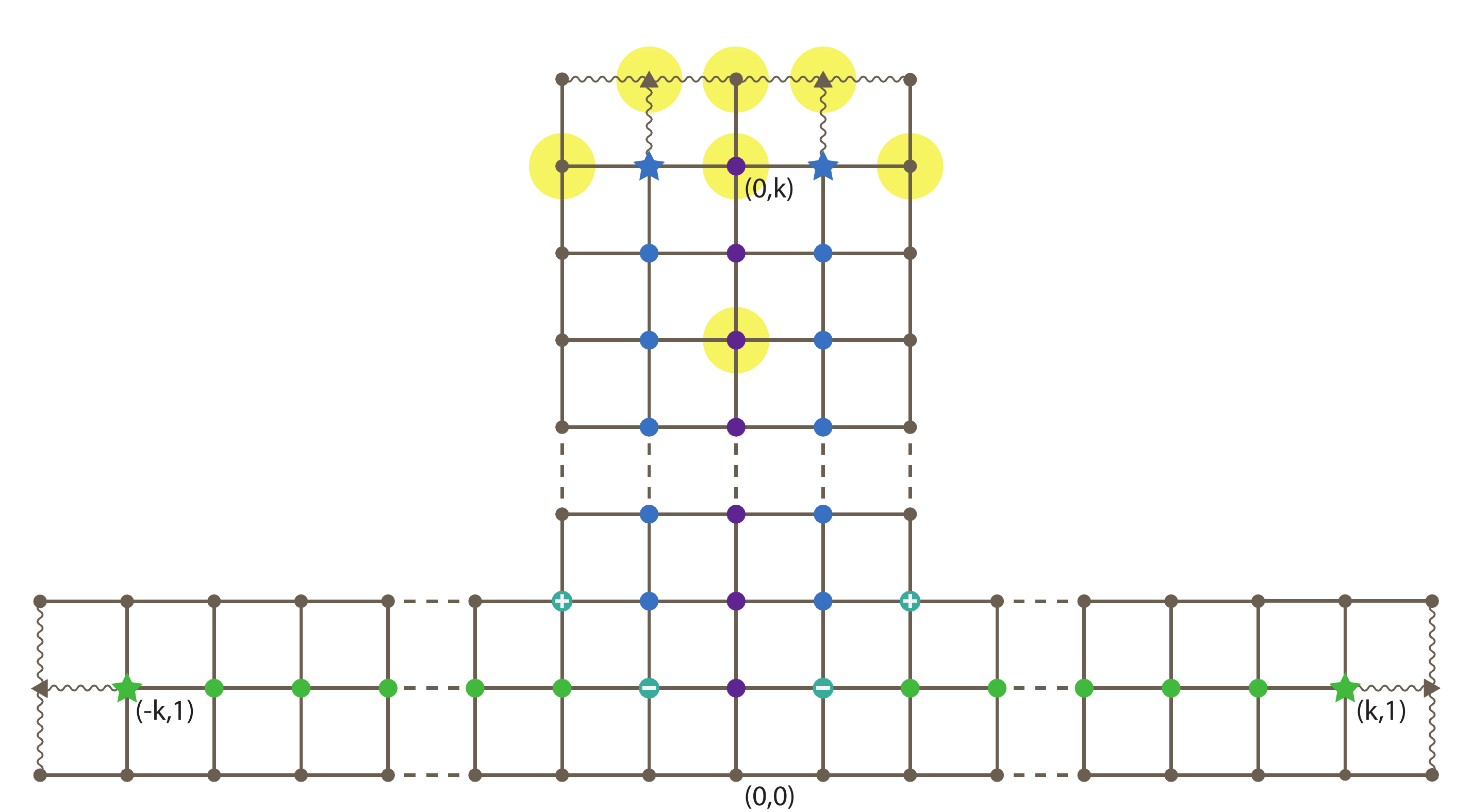}
\caption{The finite T-hook lattice for T-functions. Also here there is a T-function $T_{a,s}$ indicated by a gray dot at any lattice site $(s,a)$. The gray triangles indicate the `non-Hirota type' T-functions. By keeping these T-functions the nontrivial boundary consistency conditions are equivalent to the set of relatively simple nearest neighbour equations \eqref{eq:tildeThor}-\eqref{eq:tildeTvert}. These relations are indicated by the wavy lines, while straight lines indicate the standard Hirota relations. Equation \eqref{eq:Tk+1,0} for $T_{k+1,0}$ remains non nearest neighbour, involving the T-functions in the yellow bubbles.}
\label{fig:Ttildehook}
\end{center}
\end{figure}

Before eliminating the `non-Hirota' T-functions, the Y-system is equivalent to the bulk Hirota equations plus additional equations for $T_{1,\pm(k+1)}$, $T_{k+1,\pm1}$, and $T_{k+1,0}$. Firstly, it is easy to see that
the Y-system equations for $w$ strings for $M=k-2$ and $M=k-1$ give
\begin{align}
& T_{1, k}^+ T_{1,k}^- =
(T_{0,k}T_{2,k} + T_{1, k-1}T_{1, k+ 1})
\!\left[1+e^{\chi^{(+)}_{w}}
\frac{T_{1,k-1}T_{1,k+ 1}}{T_{0,k}T_{2,k}}\right]\!, \label{eq:tildeThor}\\
& T_{1, k+1}^+ T_{1, k+ 1}^- = e^{-\chi^{(+)}_{w}} T_{0,k+1}T_{2,k+1}\,.
\end{align}
Note the dependence on the chemical potential, and that the first factor on the left hand side of the first equation is of the usual Hirota form. Similarly, the last two equations for $vw$-strings are
\begin{align}
& T_{k, 1}^+T_{k,1}^- =
(T_{k,0}T_{k,2} + T_{k-1, 1}T_{k+1,1})
\!\left[1 + e^{\chi^{(+)}_{vw}}
\frac{T_{k-1, 1}T_{k+1,1}}{T_{k,0}T_{k,2}}\right],\\
& \tilde{T}_{k+1,1}^+\tilde{T}_{k+1, 1}^- =e^{-\chi^{(+)}_{vw}} T_{k+1,0}T_{k+1,2}\,.\label{eq:tildeTvert}
\end{align}
Of course there are similar equations for the $(-)$ wing. In addition, the equation for $Y_k$ gives
\begin{align}
T^+_{k+1,0}T^-_{k+1,0} =
\frac{T_{k+1,1}^2}{T_{k,2}}
\frac{T_{k+1,-1}^2}{T_{k,-2}}
\frac{T_{k-2,0}}{T_{k,0}}.\label{eq:Tk+1,0}
\end{align}
These equations together with the bulk Hirota equations define our T-functions on a finite - and again truly fat - hook, as illustrated in figure \ref{fig:Ttildehook}. Restricted to the auxiliary problem the functional relations satisfied by the T-functions only involve the conventional nearest neighbours. The equation for $T_{k+1,0}$ necessarily involves non nearest neighbour terms however, and as $T_{k+1,0}$ already enters in a regular Hirota equation it does not appear to make much sense to attempt to change this by introducing further auxiliary T-functions.

The parametrization of Y-functions in terms of T-functions is left invariant under
\begin{equation}
\label{eq:Tgaugetf}
T_{a,s} \rightarrow g_1^{[a+s]}g_2^{[a-s]}g_3^{[-a+s]}g_4^{[-a-s]}T_{a,s}\, ,
\end{equation}
where the $g_i$ are arbitrary functions, presenting a gauge transformation on the T-functions. Because the Y-system is invariant, the T-system is invariant under these transformations as well. This in particular includes the special equations \eqref{eq:tildeThor}-\eqref{eq:Tk+1,0} as can also be verified by direct substitution.

Asymptotically the `Hirota type' T-functions coincide with the transfer matrix based on the $\mathfrak{psu}_q (2|2)^{\oplus 2}$ invariant $S$-matrix. It should be clear however that $T_{1,\pm(k+1)},T_{k+1,\pm1}$ require some care; we will see this explicitly below. Let us now derive the transfer matrix and
then discuss how it provides us with the asymptotic solution to the T-system.

\section{Transfer matrix and fusion}\label{sec;ABA}
\label{sec:Transfermatrix}

In this section we derive the (twisted) bound state transfer matrix from first
principles. We will also show that it can be obtained by a fusion procedure. We
will work on the $u$-plane (additive parameter). In this presentation the
parameters $x^\pm$ describing a $M$-particle bound state can be conveniently
defined via $x^{\pm} = x(u\pm\frac{iM}{g})$. More generally, let us introduce
the notation $x^{[a]} \equiv x(u+\frac{ai}{g})$, then these parameters satisfy
\begin{align}
\frac{1}{q^a} \left(x^{[a]} + \frac{1}{x^{[a]}} \right) - q^a \left(x^{[-a]} +
\frac{1}{x^{[-a]}} \right) = \left(q^a - \frac{1}{q^a}\right)\left(\xi +
\frac{1}{\xi}\right)\, ,
\end{align}
where
\begin{equation}
\xi = \sqrt{\frac{g^2 \sin^2\frac{\pi}{k}}{1 + g^2 \sin^2\frac{\pi}{k}}}\, .
\end{equation}
The derivation presented here will be very similar to the ones presented in
\cite{ALST,dLvT2} and we will mainly restrict ourselves to highlighting
the differences.

\subsection{Algebraic Bethe Ansatz}

\paragraph{General procedure.} Let us first briefly outline the general
procedure. We consider $\KK{I}$
fundamental particles with rapidities $u_1,\ldots, u_{\KK{I}}$. To these
particles we add an auxiliary $a$-particle bound state with rapidity $u$. This
system is described by the following tensor product space
\begin{eqnarray}
\mathcal{V}:=V_a(u)\otimes \underbrace{V_1(u_1)\otimes \ldots \otimes
V_{\KK{I}}(u_{\KK{I}})}_{V_p},
\end{eqnarray}
where $V_i$ is the fundamental representation with rapidity $u_i$. We
divide the states into an auxiliary piece $V_a$ and a physical piece $V_p$. The
monodromy matrix is then defined as follows
\begin{eqnarray}\label{eqn;Monodromy}
\mathcal{T}_a(u ,\{u_i\}) := \prod_{i=1}^{K^{\rm{I}}}\mathbb{S}_{ai}(u,u_i).
\end{eqnarray}
This operator manifestly depends on the representation of the auxiliary particle.

The monodromy matrix is a matrix in the auxiliary space $V_a(u)$ with matrix
entries being operators acting on $V_P$. Taking a basis $|e_{I}\rangle$ (see
\eqref{eqn;basise}) for $V_a(u)$ and a basis $|f_{A}\rangle$ for $V_P$, the
action of the monodromy matrix $\mathcal{T}\equiv \mathcal{T}_a(u ,\{u_i\}) $
on the total space $\mathcal{V}$ can be written as
\begin{eqnarray}\label{eqn;ActionMonodromy}
\mathcal{T}(|e_{I}\rangle\otimes |f_A\rangle) = \sum_{J,B}
T_{IA}^{JB}|e_{J}\rangle\otimes |f_B\rangle.
\end{eqnarray}
The matrix entries of the monodromy matrix can then accordingly be denoted as
\begin{eqnarray}
\mathcal{T}|e_{I}\rangle = \sum_{J} T^{J}_{I}
|e_{J}\rangle\,  ,
\end{eqnarray}
while the action of the matrix elements $T^{J}_{I}$ as operators on $V_P$ can
easily be read off to be
\begin{eqnarray}
T^{J}_{I}|f_A\rangle = \sum_{B} T_{IA}^{JB}|f_B\rangle.
\end{eqnarray}
The operators $T^{J}_{I}$ satisfy non-trivial commutation relations among
themselves. Consider two different auxiliary spaces $V_a(u_a),V_b(u_b)$. The
Yang-Baxter equation for the S-matrix implies that
\begin{align}\label{eqn;YBE-operators}
\mathbb{S}_{ab} (u_a,u_b)\,\mathcal{T}_a(u_a ,\{u_i\}) \mathcal{T}_b(u_b
,\{u_i\})
 =
\mathcal{T}_b(u_b ,\{u_i\})\mathcal{T}_a(u_a ,\{u_i\})\,\mathbb{S}_{ab}
(u_a,u_b),
\end{align}
where $\mathbb{S}_{ab} (u_a,u_b)$ is the S-matrix describing the
scattering between the two auxiliary particles. By explicitly working out these
relations, we can find the commutation relations between the different matrix
elements of the monodromy matrix. These fundamental commutation relations
(\ref{eqn;YBE-operators}) constitute the cornerstone of the Algebraic Bethe
Ansatz \cite{FaddeevABA}.

The transfer matrix is subsequently defined as
\begin{eqnarray}
t_a(u ,\{u_i\}):={\rm{str}}_a\mathcal{T}_{a}(u ,\{u_i\}) =
(-1)^{|I|}\mathcal{T}^I_I,
\end{eqnarray}
and it can be viewed as an operator acting on the physical space $V_P$. By
virtue of the Yang-Baxter equation it defines an infinite set of commuting
quantities
\begin{align}
[t_a(u_a),t_b(u_b)] =0,
\end{align}
making the integrability of the model explicit. Our goal is to find the
eigenvalues of the transfer matrix by means of the algebraic Bethe Ansatz.

\paragraph{Twists.} Since the coproduct of the Cartan generators $H_i$ is
unaffected by quantum transformations, the group-like element $G = \exp (i\alpha
H_1 + i\beta H_3)$ gives a symmetry of the S-matrix in the sense that
\begin{align}\label{eqn;twistS}
[G_a\otimes G_b,\S_{ab}] = 0.
\end{align}
We can then introduce a family of twisted transfer matrices
\begin{align}
t^G_a(u ,\{u_i\}):={\rm{str}}_a\, G_a\mathcal{T}_{a}(u ,\{u_i\}).
\end{align}
As a consequence of \eqref{eqn;twistS}, the twisted transfer matrix preserves
the integrability property $[t^G_a(u_a),t^G_b(u_b)] =0$. Such a twisted
transfer matrix describes an integrable model with quasi-periodic boundary
conditions \cite{Sklyanin}.

\subsection{Transfer Matrix on the Vacuum}

The algebraic Bethe ansatz basically consists of two ingredients. One defines
a vacuum state and a set of creation operators that are elements of the
monodromy matrix. Eigenstates of the transfer matrix are then build by acting
with the creation operators on the vacuum. Like always, the transfer matrix
respects the two $\alg{su}_q(2)$ symmetries of the symmetry algebra of the
S-matrix. In particular, the transfer matrix commutes with the Cartan generators
$H_{1,3}$ and consequently, the eigenspaces of the transfer matrix are labelled
by the quantum labels $\KK{II},\KK{III}$ corresponding to the eigenvalues of
$H_{1,3}$. Explicitly, for a state in $V_p$ of the form\footnote{See appendix \ref{app:boundstatereps} for details and notation regarding the fundamental representation.}
\begin{align}
\bigotimes_{i=1}^{\KK{I}} |m_i,n_i,k_i,l_i\rangle,
\end{align}
we have
\begin{align}
&\KK{II} = \sum_{i=1}^{\KK{I}} m_i+n_i+2k_i, &&\KK{III} =
\sum_{i=1}^{\KK{I}} m_i+k_i.
\end{align}
These numbers once again count the total number of fermions ($\KK{II}$) and the
total number of fermions of a specific species  ($\KK{III}$).

In our case, we define the vacuum to be the highest weight state corresponding
to the quantum numbers $\KK{II}=\KK{III}=0$
\begin{align}
|0\rangle = \bigotimes_{i=1}^{\KK{I}}|0,0,0,1\rangle.
\end{align}
From the explicit form of the monodromy matrix \eqref{eqn;Monodromy} it is then
straightforwardly seen that this is an eigenstate of the transfer matrix, with
corresponding eigenvalue written in terms of S-matrix entries
\begin{align}\label{eqn;prevactrans}
\Lambda=&e^{i\alpha a} \prod_{i=1}^{\KK{I}}\mathscr{Z}^{0,0;1}_{0;1}(u,u_i) +
e^{-i\alpha a}\prod_{i=1}^{\KK{I}}\mathscr{Z}^{a,0;1}_{a;1}(u,u_i)
 - \sum_{\gamma=3}^4\sum_{m=0}^{a-1} \frac{e^{i\alpha
(a-2m-1)}}{e^{(-1)^\gamma i\beta}}\prod_{i=1}^{\KK{I}}
\mathscr{Y}^{m,0;\gamma}_{m;\gamma}(u,u_i) +\nonumber\\
&+ \sum_{a_i  = 1,3} \sum_{m=1}^{a-1}
e^{i\alpha
(a-2m)} \mathscr{Z}^{m,0;a_1}_{m;a_2}(u,u_1)\mathscr{Z}^{m,0;a_2}_{m;a_3}(u,
u_2)\ldots \mathscr{Z}^{m,0;a_{\KK{I}}}_{m;a_1} (u,u_{\KK{I}}).
\end{align}
The last term can be mapped to a $2\times 2$ matrix eigenvalue problem
just like in the undeformed case \cite{ALST} involving the eigenvalues
\begin{align}\label{eqn;lambdanice}
\lambda_\pm = \frac{x^+ - x^+_i}{x^+ - x^-_i}\,
 \frac{x^-_i - (x^{[a-2m]})^{\pm1} }{ x^+_i - (x^{[a-2m]})^{\pm1}  }
\frac{\sinh\frac{\pi g}{2k}(u - u_i +
\frac{(a-2m-1)i}{g})}{\sinh\frac{\pi g}{2k}(u - u_i + \frac{(a-1)i}{g})}.
\end{align}
From the explicit S-matrix entries listed in Appendix \ref{app;Smatrix} it is
then straightforward to arrive at the following the eigenvalue of the transfer
matrix on the vacuum
\begin{align}
t_{a,1} =&e^{i\alpha a} +
e^{-i\alpha a}\prod_{i=1}^{K^{\mathrm{I}}} q^a \frac{x^- - x^-_i}{x^+-x^-_i}
\frac{\frac{1}{x^-} - x^+_i}{\frac{1}{x^+} - x^+_i}+\sum_{m=1}^{a-1}
e^{i\alpha(a-2m)}
\left(\prod_{i=1}^{K^{\mathrm{I}}} \lambda_+ + \prod_{i=1}^{K^{\mathrm{I}}}
\lambda_-\right) \\
\nonumber
&-(e^{i\beta}+e^{-i\beta})
\sum_{m=0}^{a-1}e^{i\alpha(a-2m-1)}\prod_{i=1}^{K^{\mathrm{I}}}
\frac{q^{2m-\frac{1}{2}}}{\sqrt{\frac{x^+_i}{x^-_i}}} \frac{x^+-x^+_i}{x^+
-x^-_i} \frac{\sinh\frac{\pi g}{2k}(u - u_i +
\frac{(a-2m-1)i}{g})}{\sinh\frac{\pi g}{2k}(u - u_i +
\frac{(a-1)i}{g})}.
\end{align}
In the $q\rightarrow 1$ limit it immediately reduces to the result from
\cite{dLvT2}. The untwisted transfer matrix simply follows from setting
$\alpha,\beta=0$.

\subsection{Creation operators}

General states are constructed with the help of creation operators.
Consider a monodromy matrix where the auxiliary space is in the fundamental
representation. Let us denote the basis vectors as $ |e_i\rangle$, with
$i=1,2,3,4$. Then we define the following creation operators; two fermionic
$T^1_3,T^1_4$ and one bosonic $T^1_2$ operator. They have a clear
interpretation. By acting with $T^1_a$ on the vacuum state you create an
excitation of type $a$. More precisely, the creation operators increase the
quantum numbers of the state as listed in Table \ref{tab;creation}.

\begin{table}[h]
\begin{center}
\begin{tabular}{c|c|c|}
& $\Delta \KK{II}$ & $\Delta \KK{III}$ \\
\hline
$T^1_4$ & 1 & 1 \\
\hline
$T^1_3$ & 1 & 0 \\
\hline
$T^1_2$ & 2 & 1 \\
\hline
\end{tabular}
\caption{The quantum numbers of the creation operators.}
\end{center}
\label{tab;creation}
\end{table}

Notice that any state with quantum numbers $\KK{II},\KK{III}$ can be
constructed by applying these generators. Moreover, we see that both
$T^1_3T^1_4$ and $T^1_2$ have the same quantum numbers, which indicates that
they will mix. This has its origin in the presence of the scattering process
that mixes bosons and fermions. In other words, a generic state will be a
linear combination of all the operators that create the same quantum numbers. Of
course, this can only be an eigenstate for specific values of the relative
coefficients.

The last crucial ingredient that is needed are the commutation relations. By
computing the commutation relation between two creation operators, a nested
structure is revealed. The fermionic creation operators can be commuted via an
$\alg{su}(2)$ invariant S-matrix, $r$, in the following way
\begin{align}
T^1_\alpha(u_1) T^1_\beta(u_2) =& -\frac{x_1^- - x_2^+}{x_1^+ -
x_2^-}\frac{U_1V_1}{U_2V_2} \, r^{\gamma\delta}_{\alpha\beta}(u_1-u_2)\,
T^1_\delta(u_2) T^1_\gamma(u_1)  + \\
&+ \frac{\epsilon_{\alpha\beta}}{q^{\epsilon_{\alpha\beta}}}
\frac{(x^+_1-x^-_1)(x^+_2-x^-_2)}{1- x^+_1 x^-_2}
\frac{U_1V_1}{\gamma_1\gamma_2} (T^1_2(u_2) T^1_1(u_1) - T^1_2(u_1)
T^1_1(u_2)),\nonumber
\end{align}
where  $\alpha,\beta=3,4$ and we take $\epsilon_{34} = 1$. The non-zero
components of $r(u)$ are $r_{33}^{33} = r_{44}^{44} =1$ and
\begin{align}\label{eqn;auxr}
&r^{34}_{43}  = r^{43}_{34}  =
\frac{\sinh\frac{g\pi u}{2k}}{\sinh\frac{g\pi}{2k}(u-\frac{2i}{g}
)}, && r_{34}^{34} =
\frac{e^{\frac{g\pi u}{2k}}\sin\frac{\pi}{k}}{\sinh\frac{g\pi}{2k}
(u-\frac{2i}{g})}, &&
r_{43}^{43} =
\frac{e^{-\frac{g\pi u}{2k}}\sin\frac{\pi}{k}}{\sinh\frac{g\pi}{2k}
(u-\frac{2i}{g})}.
\end{align}
and it is simply the S-matrix of an inhomogeneous XXZ spin chain. As such it is
unitary and satisfies the Yang-Baxter equation.

The relevant parts of the other commutation relations that deal with the
diagonal components of the monodromy matrix, and hence are important for the
transfer matrix, are listed in Appendix \ref{app;commrel}.

\subsection{Full Transfer Matrix}

In \cite{ALST} the eigenvalues of the transfer matrix were computed quite
generally in terms of elements of the S-matrix in (3.42). Indeed, it applies
here as well, but the auxiliary problem is now described by \eqref{eqn;auxr}
rather than the XXX spin chain. Let us introduce a short-hand notation for the
different types of functions that appear
\begin{align}
&\SH^A_B =  \prod_{i=1}^{K^A}\sinh\frac{\pi g}{2k}(u-u_i ^{(A)} + B\frac{i}{g}),
&& R^\pm_A = \prod_{n=1}^{K^{\mathrm{I}}}  \frac{x(u+\frac{Ai}{g}) - x^{\mp}_{n}
}{q^\frac{A\mp1}{4}\sqrt{x^{\mp}_{n} }} ,
&& B^\pm_A = \prod_{n=1}^{K^{\mathrm{I}}}  \frac{\frac{1}{x(u+\frac{Ai}{g})} -
x^{\mp}_{n}}{q^\frac{A\mp1}{4}\sqrt{x^{\mp}_{n} }} ,
\end{align}
where $u^{(\mathrm{II})}_i =v_i$ and
$u^{(\mathrm{III})}_i = w_i$. We then arrive at
\begin{align}\label{fulltransfer}
t_{a,1}=& e^{i\alpha
a}\frac{\SH^\mathrm{II}_a}{R^+_aB^-_a} \prod_{i=1}^{K^{\mathrm{II}}}
q^{\frac{a}{2}}\frac{y_i - x^-}{y_i
-x^+}\sqrt{\frac{x^+}{x^-}}\left\{\sum_{m=0}^{a}
\frac{R^{+}_{a-2m}B^{-}_{a-2m}}{e^{2i\alpha m}\,\SH^\mathrm{II}_{a-2m}}
+
\sum_{m=1}^{a-1}
\frac{R^{-}_{a-2m}B^{+}_{a-2m}}{e^{2i\alpha m}\,\SH^\mathrm{II}_{a-2m}}
\right.\nonumber\\
& - \sum_{m=0}^{a-1}\!
\frac{R^{-}_{a-2m}B^{-}_{a-2m}}{e^{i\alpha(2m+1)}\SH^\mathrm{III}_{a-2m-1}}
\left.
\left[\!
e^{i\beta}
\frac{\SH^\mathrm{III}_{a-2m+1}}{\SH^\mathrm{II}_{a-2m}}
+e^{-i\beta}
\frac{\SH^\mathrm{III}_{a-2m-3}}{\SH^\mathrm{II}_{a-2m-2}}
\right]
\right\},
\end{align}
while the auxiliary parameters $y_m$ and $w_m$ satisfy the auxiliary Bethe
equations
\begin{align}
\label{eq:auxyfinal}
1&= e^{i(\alpha-\beta)}\prod_{i=1}^{K^{\mathrm{I}}}
\sqrt{q}\frac{y_m -x^-_i}{y_m - x^+_i}\sqrt{\frac{x^+_i}{x^-_i}}
\prod_{i=1}^{K^{\mathrm{III}}}
\frac{\sinh{\frac{\pi g}{2k}\big(v_m - w_i-\frac{i}{g}\big)}}
{\sinh{\frac{\pi g}{2k}\big(v_m - w_i+\frac{i}{g}\big)}}\, ,\\
\label{eq:auxwfinal}
-1&= e^{-2i\beta}\prod_{i=1}^{K^{\mathrm{II}}} \frac{\sinh{\frac{\pi
g}{2k}\big(w_n - v_i + \frac{i}{g}\big)}}{\sinh{\frac{\pi
g}{2k}\big(w_n - v_i -
\frac{i}{g}\big)}}\prod_{j=1}^{K^{\mathrm{III}}}\frac{\sinh{\frac{\pi
g}{2k}\big(w_n - w_j - \frac{2i}{g}\big)}}{\sinh{\frac{\pi
g}{2k}\big(w_n - w_j + \frac{2i}{g}\big)}}\, ,
\end{align}
where
\begin{equation}
v_j = -\frac{y_j +\frac{1}{y_j} + \xi + \frac{1}{\xi}}{\xi - \frac{1}{\xi}}
\end{equation}
According to formalism of the analytic Bethe Ansatz these follow from the pole
structure of the transfer matrix. The main Bethe equations pick up a
a twist factor $e^{i\alpha}$ as in \cite{dLvT2}.

\subsection{Generating Functional}

There is an alternative way to derive the transfer matrix in higher
representations from the fundamental one. This method goes under the name of
generating functional \cite{Kirillov:1987zz,Krichever:1996qd,Tsuboi:1998ne,Beisert:2006qh}.

Up to an overall normalization, we can write the fundamental transfer matrix as
a sum of four terms
\begin{align}
t_{1,1} = e^{i\alpha} +
e^{-i\alpha}\frac{R^+_{-1} B^-_{-1} \SH^\mathrm{II}_1}{R^+_{1} B^-_{1}
\SH^\mathrm{II}_{-1}} -
e^{i\beta}\frac{R^-_{1} \SH^\mathrm{III}_2}{R^+_{1} \SH^\mathrm{III}_0} -
e^{-i\beta}\frac{R^-_{1} \SH^\mathrm{II}_1 \SH^\mathrm{III}_{-2}}{R^+_{1}
\SH^\mathrm{II}_{-1} \SH^\mathrm{III}_0}
\equiv t_1 + t_2 - t_3 - t_4.
\end{align}
Consider the shift operator $D$ which is defined via  $D f(u) =
f(u-\frac{i}{g})D$, then we can introduce the generating functional $W$
\begin{align}
W =
\frac{1}{1 - D\, t_2\, D}
\left[1- D\, t_4\, D\right]
\left[1-D\, t_3\, D\right]
\frac{1}{1- D\, t_1\, D}.
\end{align}
The above defined functional $W$ then generates the different transfer matrices
by expanding in the shift operator $D$ according to
\begin{align}\label{eqn;genW}
& W = \sum_{a}D^a t_{a,1}D^a, && W^{-1} = \sum_{s}D^s t_{1,s}D^s\,.
\end{align}
The fundamental transfer matrix agrees by
construction. It is then straightforward, but tedious, to match the transfer
matrix term by term.

Let us also spell out the transfer matrix $t_{1,s}$, which can be computed via
the inverse generating functional according to \eqref{eqn;genW}. This transfer
matrix will prove to be useful in defining some Y-functions later. The transfer
matrix is generically defined up to normalization, so let us adapt a convenient
one for this purpose by setting
\begin{align}
t_3\rightarrow e^{i\beta}.
\end{align}
Then it is straightforward to work out $W^{-1}$ explicitly
\begin{align}\label{eqn;Ts}
t_{1,s}  =&
\frac{\SH^\mathrm{III}_{1-s}}{e^{i\beta s}\SH^\mathrm{II}_{-s}}\left[
\sum_{m=0}^{s}
\frac{e^{2 i\beta m}\SH^\mathrm{II}_{s-2m}\SH^\mathrm{III}_{-1-s}}
{\SH^\mathrm{III}_{s-2m-1}\SH^\mathrm{III}_{s-2m+1}}
- e^{i\alpha}\frac{R^+_s}{R^-_s}
\sum_{m=1}^{s}
\frac{e^{i\beta (2m-1)}
\SH^\mathrm{II}_{s-2m}\SH^\mathrm{III}_{-1-s}\SH^\mathrm{III}_{s-1}}
{\SH^\mathrm{III}_{s-2m-1}\SH^\mathrm{III}_{s-2m+1}\SH^\mathrm{III}_{s+1}}
+\right.
\\
&\left.
\frac{R^+_s B^-_{-s}}{R^-_s B^+_{-s}}
\sum_{m=1}^{s-1}
\frac{e^{2i\beta m}
\SH^\mathrm{II}_{s-2m}\SH^\mathrm{III}_{1-s}\SH^\mathrm{III}_{ s-1}}
{\SH^\mathrm{III}_{s-2m-1}\SH^\mathrm{III}_{s-2m+1}\SH^\mathrm{III}_{s+1}}
-e^{-i\alpha}\frac{B^-_{-s}}{B^+_{-s}}
\sum_{m=0}^{s-1}
\frac{e^{i\beta (2m+1)}\SH^\mathrm{II}_{s-2m}\SH^\mathrm{III}_{1-s}}
{\SH^\mathrm{III}_{s-2m-1}\SH^\mathrm{III}_{s-2m+1}}
\right],\nonumber
\end{align}
where we used that $R^+_{-s}B^+_{-s}=R^-_{s}B^-_{s}$. A different, related
procedure is that of dualizing the transfer matrix.
Details are provided in Appendix \ref{app;dual} where we compare $t_{1,s}$
to the dualized version of $t_{a,1}$.

\section{Asymptotic solution}
\label{sec:Asymptoticsolution}

Now that we have derived the twisted transfer matrix, we can construct a
solution for the Y-system when $Y_Q\rightarrow Y^\circ_Q\sim 0$. The Y-system in
this limit is called the asymptotic Y-system, as it arises for example when $J$
is large and the asymptotic Bethe Ansatz should describe the problem exactly. In
this section all Y-functions will be understood in the asymptotic limit and to
avoid cluttered notation, we will simply identify all Y-functions with their asymptotic
limit and drop the superscript $\circ$.

\subsection{Asymptotic Y-system}

Let us first summarize the asymptotic Y-system equations for the different types
of strings. As before, we only label the left and right copies of $\alg{su}_q(2|2)$ explicitly in the equations for $Q$-particles.

\paragraph{$w$ strings.} The equations for $w$ strings do not contain the $Y_Q$
functions so their form is simply unaffected
\begin{align}
Y_{1|w}^+ Y_{1|w}^- & =(1+Y_{2|w})\left(\frac{1-Y_-^{-1}}{1-Y_+^{-1}}\right)^{\theta(u_B-|u|)}  \\
Y_{M|w}^+ Y_{M|w}^- & =(1+Y_{M+1|w})(1+Y_{M-1|w}) \, \, , \, \, \, M=2,\ldots,k-3 \, ,\\
Y_{k-2|w}^+ Y_{k-2|w}^- & = (1+Y_{k-3|w})(1+Y_{k-1|w})(1 + e^{\chi_{w}} Y_{k-1|w}) \, ,\label{eqn;Zw1}\\
Y_{k-1|w}^+ Y_{k-1|w}^- & = e^{-\chi_{w}}(1+Y_{k-2|w}) \, .\label{eqn;Zw2}
\end{align}
\paragraph{$vw$ strings.} By dropping the $Y_Q$ terms we get a similar set of
equations for $vw$-strings
\begin{align}
Y_{1|vw}^+ Y_{1|vw}^- & =(1+Y_{2|vw})\left(\frac{1-Y_-}{1-Y_+}\right)^{\theta(u_B-|u|)}  \\
Y_{M|vw}^+ Y_{M|vw}^- & =(1+Y_{M+1|vw})(1+Y_{M-1|vw})\, ,\\
Y_{k-2|vw}^+ Y_{k-2|vw}^- & = (1+Y_{k-3|vw})(1 + Y_{k-1|vw})
(1 +e^{\chi_{vw}} Y_{k-1|vw})\, ,\\
Y_{k-1|vw}^+ Y_{k-1|vw}^- & = e^{-\chi_{vw}}(1+Y_{k-2|vw})\, .
\end{align}
Apart from the first equation they are exactly of the same form as those for $w$ strings.

\paragraph{$y$ strings.} For $y$ particles there is only one local equation. Asymptotically it reduces to
\begin{equation}
Y_{-}^{+}Y_{-}^{-} = \frac{1+Y_{1|vw}}{1+Y_{1|w}}\, .
\end{equation}

\paragraph{Q particles.} The Y-system for $Q$-particles involves ratios of $Y_Q$ functions. These survive in the asymptotic limit, resulting in the following set of equations. For $Q=1$ we have
\begin{align}
&\frac{Y_1^+ Y_1^-}{Y_2} = \prod_{r=\pm}
\left(1-\frac{1}{Y^{(r)}_{-}}\right)\, , \, \, \, \, \, \mbox{for} \, \,
|u|<u_b\, .
\end{align}
Then for $Q=2,\ldots,k-1$ we find
\begin{align}
\frac{Y_Q^+ Y_Q^-}{Y_{Q+1}Y_{Q-1}} = \prod_{r=\pm} \left(
1+\frac{1}{Y_{Q-1|vw}^{(r)}}\right) \, ,
\end{align}
while at the boundary $Q=k$
\begin{align}\label{eqn;Ykeqn}
&\frac{Y_k^+ Y_k^-}{Y_{k-1}^2} =\! \prod_{r=\pm}
\left(1 +\frac{1}{Y_{k-1|vw}^{(r)}}\right)
\left(1 +\frac{e^{-\chi^{(r)}_{vw}}}{Y_{k-1|vw}^{(r)}}\right).
\end{align}

\subsection{Solution via transfer matrices}

In the asymptotic regime we can find a solution of the Y-system
in terms of transfer matrices. The T-system indicates how this is done. We
started out with a model with two copies of $\alg{su}_q(2|2)$ symmetry.
These copies form the left and right wing of the fat-hook and this structure
should be manifest asymptotically. Hence, we are able to express the
T-functions in the left and right hook in terms of the $\alg{su}_q(2|2)$
transfer matrices that were derived in the previous section. Of course, the left
and right copies can in general have different twists $\alpha_\pm,\beta_\pm$.

The T-functions are defined up to gauge transformations \eqref{eq:Tgaugetf}, which we
can use to set $T_{0,s} = 1$. Then, for $a\leq k$ we make the following
identification
\begin{align}
&T_{a,\pm1} \rightarrow
\left.\tau_{a,1}t_{a,1}\right|_{\alpha=\alpha_\pm, \beta=\beta_\pm},
&& T_{a,0} \rightarrow 1,
\end{align}
where $\tau_{a,1}$ is simply a normalization of the transfer matrix such that
\begin{align}
\tau^2_{Q,1}(u) =
e^{-J\tilde{\mathcal{E}}_Q}\prod_{i=1}^{\KK{I}}
S^{Q 1_*}_{\alg{sl}(2)}(u,u_i) \equiv \Upsilon_Q \, ,
\end{align}
where the $u_i$ are the asymptotic rapidities of the excitations under consideration,
\begin{equation}
S^{Q 1_*}_{\alg{sl}(2)} = (S^{Q1})^{-1} (\Sigma^{Q 1_*})^{-2}\, ,
\end{equation}
and $\Sigma^{Q 1_*}$ is the improved mirror bound state dressing phase with its second leg continued to the `string' line; see appendix \ref{app:dressing phase} for an explicit expression for and a discussion of the properties of $\Sigma$.
The reason for this factor is that the transfer matrix \eqref{fulltransfer} is constructed from the canonically normalized
S-matrix; the $S_{\alg{sl}(2)}$ part of
$\Upsilon_Q$ simply reintroduces the correct dressing for the S-matrix. Secondly
we would like to point out that $e^{-J\tilde{\mathcal{E}}_Q}$ satisfies
the discrete Laplace equation, so it is basically a gauge transformation. We see
that this factor implies
\begin{align}
Y_Q \sim e^{-J\tilde{\mathcal{E}}_Q},
\end{align}
which makes explicit that $Y_Q$ is small for large $J$. Moreover, we now find that the asymptotic Bethe equations are equivalent to
\begin{align}
Y_{1*}(u_i) = -1\, ,
\end{align}
as in the undeformed model.

\paragraph{Bazhanov-Reshetikhin formula.} In section \ref{sec;ABA} we have explicitly constructed
the transfer matrix $t_{a,1}$ for short representations corresponding to a
Young diagram with $a$ rows and one column. However, \eqref{eqn;Tsystem} also
contains T-functions with different indices. There is a well-known
determinant formula that provides a formal solution of the Hirota equation in terms of $t_{a,1}$ as
\begin{align}\label{eqn;BR}
t_{a,s}(u) = \det_{1\leq m,n\leq s} t_{a+m-n,1}(u +i(s+1-m-n)/g) .
\end{align}
Equivalently, we could also use $t_{1,s}$ to generate all required
transfer matrices.\footnote{We remind the reader that $t_{1,s}$ in eqn. \eqref{eqn;Ts} is derived with a differently normalization from $t_{a,1}$ in eqn. \eqref{fulltransfer}. The Bazhanov-Reshetikhin formula allows us to express $t_{1,s}$ in terms of $t_{a,1}$ and doing so we find agreement with
\eqref{eqn;Ts} up to this normalization factor.}

\paragraph{Asymptotic solution.} It is now straightforward to derive the
solution of the asymptotic Y-system in terms of transfer matrices. We get the
following identification for $M<k-1$ and
$Q=1,\ldots,k$
\begin{align}\label{eqn;asymptsol}
&Y_{M|w} = \frac{t_{1,M}t_{1,M+2}}{t_{2,M+1}}, &&Y_{M|vw} =
\frac{t_{M,1}t_{M+2,1}}{t_{M+1,2,}},
&& Y_{Q} = \Upsilon_Q\, t_{Q,1} t_{Q,-1}\nonumber\\
& Y_- = \frac{t_{2,1}}{t_{1,2}} , && Y_+ = -
\frac{t_{2,3}t_{2,1}}{t_{3,2}t_{1,2}}.
\end{align}
As always, the auxiliary space uses the mirror parameterization, while the
physical particles use the `string' variables.

However, this standard construction clearly cannot hold at $M=k-1$ since
we know that $T_{1,k+1}$ and $T_{k+1,1}$ satisfy a special
set of equations. To correctly take the contributions of these functions into
account, let us set\footnote{In \eqref{eqn;asympYkvw} $t_{1,s}$ is
given by \eqref{eqn;Ts}.}
\begin{align}\label{eqn;asympYkvw}
& Y_{k-1|w} = \frac{t_{1,k-1}}{\tilde{t}_{1,k-1}},
&& Y_{k-1|vw} = \frac{t_{k-1,1}}{\tilde{t}_{k-1,1}}
\end{align}
where
\begin{align}
\tilde{t}_{1,s}  =&\frac{\SH^\mathrm{III}_{1-s}}{e^{i\beta
s}\SH^\mathrm{II}_{-s}}\left[
- e^{i\alpha}\frac{R^+_a}{R^-_a}
\sum_{m=s+1}^{s+1}
\frac{e^{i\beta (2m-1)}
\SH^\mathrm{II}_{s-2m}\SH^\mathrm{III}_{-1-s}\SH^\mathrm{III}_{s-1}}
{\SH^\mathrm{III}_{s-2m-1}\SH^\mathrm{III}_{s-2m+1}\SH^\mathrm{III}_{s+1}}
+\right.
\\
&\left.
\frac{R^+_a B^-_{-a}}{R^-_a B^+_{-a}}
\sum_{m=s}^{s+1}
\frac{e^{2i\beta m}
\SH^\mathrm{II}_{s-2m}\SH^\mathrm{III}_{1-s}\SH^\mathrm{III}_{ s-1}}
{\SH^\mathrm{III}_{s-2m-1}\SH^\mathrm{III}_{s-2m+1}\SH^\mathrm{III}_{s+1}}
-e^{-i\alpha}\frac{B^-_{-a}}{B^+_{-a}}
\sum_{m=s}^{s}
\frac{e^{i\beta (2m+1)}\SH^\mathrm{II}_{s-2m}\SH^\mathrm{III}_{1-s}}
{\SH^\mathrm{III}_{s-2m-1}\SH^\mathrm{III}_{s-2m+1}}
\right],\nonumber
\end{align}
and
\begin{align}\label{fulltransferB}
\tilde{t}_{a,1}=& e^{i\alpha
(3a+2)}\frac{\SH^\mathrm{II}_a}{R^+_aB^-_a} \prod_{i=1}^{K^{\mathrm{II}}}
q^{\frac{a}{2}}\frac{y_i - x^-}{y_i
-x^+}\sqrt{\frac{x^+}{x^-}}\left\{\sum_{m=a+1}^{a+1}
\frac{R^{+}_{a-2m}B^{-}_{a-2m}}{e^{2i\alpha m}\,\SH^\mathrm{II}_{a-2m}}
+
\sum_{m=a}^{a}
\frac{R^{-}_{a-2m}B^{+}_{a-2m}}{e^{2i\alpha m}\,\SH^\mathrm{II}_{a-2m}}
\right.\nonumber\\
& - \sum_{m=a}^{a}\!
\frac{R^{-}_{a-2m}B^{-}_{a-2m}}{e^{i\alpha(2m+1)}\SH^\mathrm{III}_{a-2m-1}}
\left.
\left[\!
e^{i\beta}
\frac{\SH^\mathrm{III}_{a-2m+1}}{\SH^\mathrm{II}_{a-2m}}
+e^{-i\beta}
\frac{\SH^\mathrm{III}_{a-2m-3}}{\SH^\mathrm{II}_{a-2m-2}}
\right]
\right\},
\end{align}
Both terms contain only four terms and as such appear to be some special case
of the fundamental transfer matrix. They are obtained by extending the sums that
appear in the general expressions \eqref{fulltransfer} and \eqref{eqn;Ts}.

From the explicit expressions for the transfer matrix, it is
straightforward to check that \eqref{eqn;asymptsol} and \eqref{eqn;asympYkvw}
indeed provide a solution to the asymptotic Y-system. Let us highlight two
reasons why this is quite non-trivial.

First, the function $Y_{k-1|(v)w}$ a priori satisfies two different equations
which relate $\tilde{t}$ to $t$, and it is not obvious that both equations are
compatible. Their compatibility crucially depends on the fact that $q^k=-1$.

The second non-trivial consistency check lies in the equation for $Y_k$. From
the asymptotic equation for $Y_{k-1}$ we can immediately deduce that $Y_k$ is in
fact given by the bulk expression at $Q=k$, while it satisfies an equation of a
different type itself. The fact that equation \eqref{eqn;Ykeqn} is satisfied
nonetheless relies crucially on the factor $\Upsilon_Q$. Indeed, since
$\Upsilon_Q$ satsifies the discrete Laplace equation
\begin{align}
\Upsilon_Q^+\Upsilon_Q^- = \Upsilon_{Q+1}\Upsilon_{Q-1},
\end{align}
we see that in the RHS of \eqref{eqn;Ykeqn} the dependence of $\Upsilon_Q$ does
not drop out. Following the arguments in Appendix E of
\cite{Arutyunov:2012zt} we can derive the following
identity
\begin{align}
\frac{\Upsilon_k^+\Upsilon_k^-}{\Upsilon_{k-1}^2} = (S^{10}_{xy})^2 =
\left[\prod^{\KK{I}}_{i=1}q\frac{x^{[k+1]}-x^-_i}{x^{[k+1]}-x^+_i}
\frac{1-\frac{1}{x^{[k-1]}x^-_i}}{1-\frac{1}{x^{[k-1]}x^+_i}}\right]^2.
\end{align}
This non-trivial factor makes \eqref{eqn;Ykeqn} hold.

\subsection{The twisted ground state solution}

As shown in \cite{Arutyunov:2012zt}, the ground state TBA equations of the deformed model have almost the same solution as the undeformed model \cite{Frolov:2009in}; only the boundary Y-functions are different. Interestingly enough it turns out that this feature persists for the twisted ground state.

\begin{figure}[h]
\begin{center}
\includegraphics[width=4in]{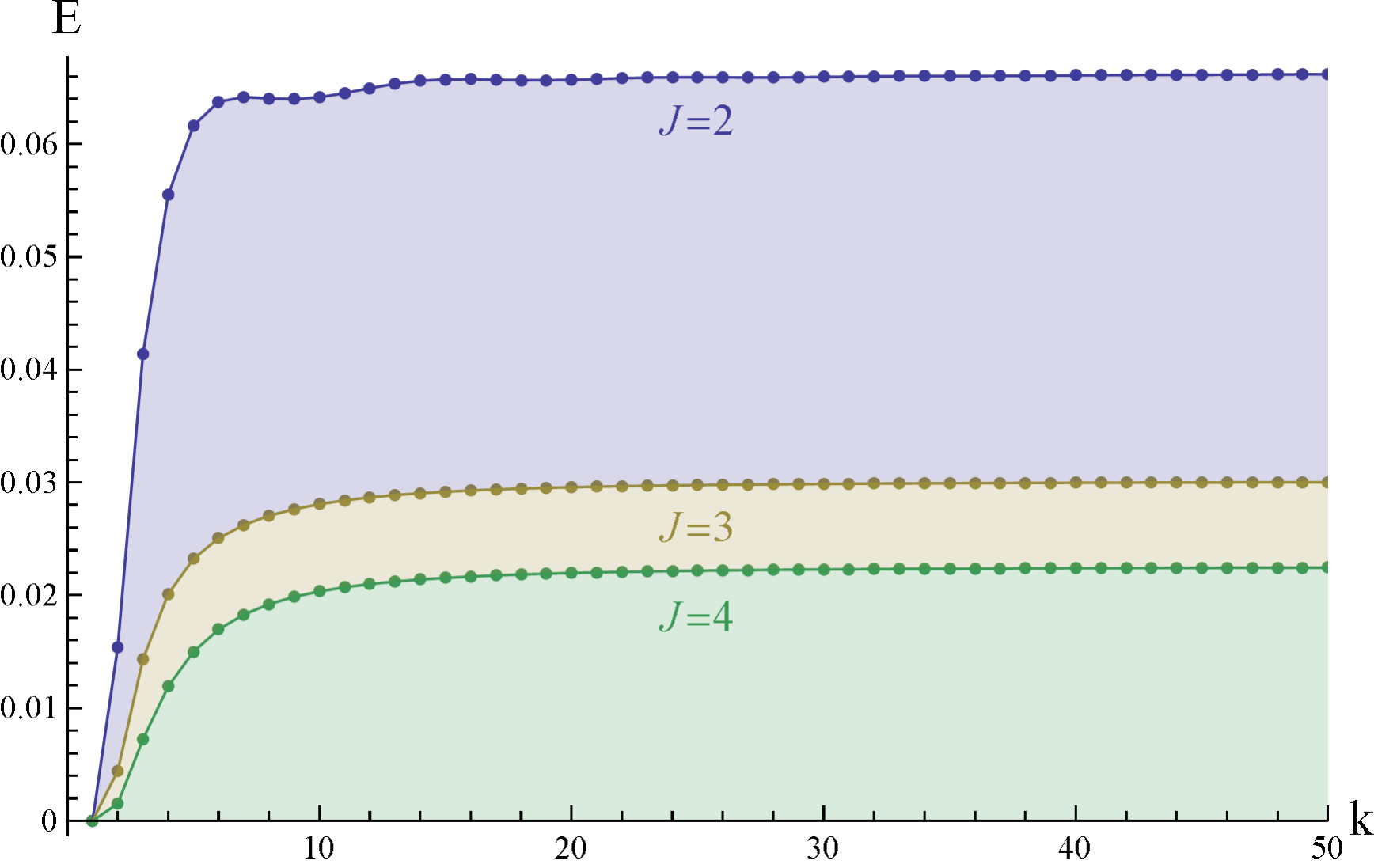}
\caption{The ground state energy of the twisted deformed theory. The plot shows the energy as a function of $k$, for $J=2,3,4$ with arbitrarily chosen twists $\alpha_+ = \tfrac{3}{10}$, $\alpha_- = \tfrac{4}{10}$, $\beta_+ = \tfrac{12}{10}$, and $\beta_- = -\tfrac{2}{10}$. The energies at $k=1$ and $k=2$ are the continuation of our results valid for $k>2$.}
\label{fig:twisteddeformedGSE}
\end{center}
\end{figure}

In the asymptotic limit, the constant bulk Y-system for the auxiliary problem is given by
\begin{align}
Y_+^2 = Y_-^2 & = \frac{1+Y_{1|vw}}{1+Y_{1|w}}\, ,\\
Y_{M|w}^2 & = (1+Y_{M-1|w})(1+Y_{M+1|w})\, ,\\
Y_{M|vw}^2 & = (1+Y_{M-1|vw})(1+Y_{M+1|vw})\, ,
\end{align}
where potential $M=0$ terms on the right hand side are zero. The general solution to these recurrence relations for the bulk $Y_{M|w}$ and $Y_{M|vw}$ is given by
\begin{align}
\label{eq:gssolnbasic}
Y_{M|w}& = [M]_{e^{i b}} [M+2]_{e^{i b}}\, , \\
Y_{M|vw} & = [M]_{e^{i a}} [M+2]_{e^{i a}} \, ,
\end{align}
where $a$ and $b$ are undetermined complex numbers for the moment, and $[x]_q = (q^{x}-q^{-x})/(q-q^{-1})$. Of course, we have yet to take into account the nontrivial boundary of the Y-system. These boundary equations read
\begin{align}
\left([k-2]_{e^{i b}} [k]_{e^{i b}}\right)^2& = (1+[k-3]_{e^{i b}}[k-1]_{e^{i b}})(1+Y_{k-1|w})(1+e^{k \mu_{1|w}}Y_{k-1|w})\, , \\
Y_{k-1|w}^2& = e^{-k \mu_{1|w}}(1+[k-2]_{e^{i b}}[k]_{e^{i b}}) \, ,
\end{align}
with a similar equation for $Y_{k-1|vw}$. These equations do not have a unique solution for arbitrary $\mu_{1|(v)w}$ and $b (a)$. However, insisting that these equations satisfy the canonical TBA equations we immediately find that $a$ and $b$ are precisely $\alpha$ and $\beta$ respectively as in table \ref{tab:chempot}, and we get
\begin{align}
\label{eq:gssolnbndry}
Y_{k-1|w} & = e^{-i k \beta} [k-1]_{e^{i\beta}}\, ,\\
Y_{k-1|vw} & = e^{-i k \alpha} [k-1]_{e^{i\alpha}}\, .
\end{align}
With this solution we know $Y_\pm$ by the Y-system equation above and can then find the $Y_Q$ functions by their canonical TBA equations. The resulting $Y_Q$ functions are formally identical to the $Y_Q$ functions of the undeformed twisted ground state and are given by
\begin{equation}
\label{eq:gssolnYQ}
Y_{Q} =  ([2]_{e^{i\alpha_+}}-[2]_{e^{i\beta_+}})([2]_{e^{i\alpha_-}}-[2]_{e^{i\beta_-}})[Q]_{e^{i\alpha_+}}[Q]_{e^{i\alpha_-}} e^{-J \tilde{\mathcal{E}}_Q(\tilde{p})}\, ,
\end{equation}
As the entire solution except the boundary $w$ and $vw$ Y-functions formally coincides with the undeformed twisted ground state solution \cite{arXiv:1108.4914,dLvT2}, this solution of the deformed TBA equations smoothly turns into the twisted ground state solution of the undeformed model in the simplest way possible. Computing the ground state energy for these solutions at leading order in $g$ gives complicated expressions; instead of giving them we have plotted the energies for some generic parameter values in figure \ref{fig:twisteddeformedGSE}.
The energy converges to the undeformed result as $\tfrac{1}{k^2}$ at large $k$. Explicitly
\begin{equation}
 E(k,J) = E(J) - \frac{J\pi^2}{3k^2} \left(E(J) - \frac{J-2}{4(J-1)} E(J-1)\right) + \mathcal{O}\left(\frac{1}{k^3}\right)\, ,
\end{equation}
where $E(J)$ is the ground state energy of the undeformed twisted model \cite{dLvT2}.

\section{The relativistic limit}
\label{sec:Relativisticlimit}

Keeping $k$ fixed and taking $g\rightarrow \infty$ in an appropriate fashion, we obtain a set of TBA equations based on the conjectured S-matrix of the Pohlmeyer reduced superstring, a semi-symmetric space sine-Gordon (ssssG) theory. The appropriate limit to take is to rescale our rapidities $u \rightarrow \frac{\tilde{u}}{g}$ and take $g \rightarrow \infty$ keeping $\tilde{u}$ fixed. In what follows by conventional abuse of notation we drop the tilde. In this limit the full S-matrix and consequently all S-matrices entering the Bethe-Yang equations become of difference form as appropriate for a relativistic theory.

At the level of the simplified TBA equations this relativistic limit is implemented in a very simple fashion. In the interpolating theory have the following three types of convolutions
\begin{align}
f\star s(u,v)=\,&\int_{-\infty}^{\infty}\, dt\, f(u,t)s(t-v) \, , \label{eq:stdconv}\\
f\, \hat{\star}\,  s(u,v)=\,&\int_{-u_b}^{u_b}\, dt\, f(u,t)s(t-v)\, , \label{eq:hatconv}\\
f\, \check{\star}\,  s(u,v)=\,&\int_{-\infty}^{-u_b}\, dt\, f(u,t)s(t-v)+\int_{u_b}^{\infty}\, dt\, f(u,t)s(t-v)\label{eq:checkconv}\, .
\end{align}
where
\begin{equation}
s(u)=\frac{g}{4 \cosh\frac{g \pi u}{2}}\, ,
\end{equation}
and
\begin{equation}
u_b = \frac{k}{\pi g} \log\frac{1+\xi}{1-\xi}=\frac{2k}{\pi g}{\rm arcsinh}\Big(g\sin\frac{\pi}{k}\Big)\, .
\end{equation}
Rescaling the rapidities and taking the infinite coupling limit the points $\pm \, g \, u_b$ go to positive and negative infinity respectively, and consequently the limit can be summarized as
\begin{align}
f\star s & \rightarrow f\star s \, , \\
f\, \hat{\star}\, s & \rightarrow f\star s\, , \\
f\, \check{\star}\,  s& \rightarrow 0\, ,
\end{align}
where on the right hand side we of course have the properly rescaled kernel
\begin{equation}
s(u) = \frac{1}{4 \cosh\frac{\pi u}{2}}\, .
\end{equation}
As a quick example of this simple consideration, let us consider the mirror energy satisfying the following identity
\begin{equation}
\tilde{\mathcal{E}}_M (K+1)^{-1}_{MQ} = \delta_{Q,1} \check{\tilde{\mathcal{E}}} \check{\star} s \, .
\end{equation}
This implies that in the relativistic limit we should have
\begin{equation}
\tilde{\mathcal{E}}_1 - \tilde{\mathcal{E}}_2 \star s = 0 \, ,
\end{equation}
where in this limit
\begin{equation}
\lim_{g\rightarrow\infty} \, g \,\tilde{\mathcal{E}}_Q\left(\tfrac{u}{g}\right) = 2 \cosh \frac{\pi u }{2k} \, \frac{\sin \frac{\pi Q}{2k}}{\sin \frac{\pi}{k}} \, .
\end{equation}
Since $\tilde{\mathcal{E}}_Q$ is a meromorphic function on the whole $u$-plane in this limit, the identity above nicely follows from
\begin{equation}
\tilde{\mathcal{E}}_1^+ +  \tilde{\mathcal{E}}_1^-  = \tilde{\mathcal{E}}_2 \, .
\end{equation}

In this limit the discontinuities of the Y-functions following from the TBA equations have disappeared completely. The dressing phase in particular only enters in defining the asymptotics of the Y-functions but does not enter in the TBA equations explicitly anymore. This is because before taking the relativistic limit the dressing phase enters the simplified TBA equations with a $\check{\star}$ convolution, as we have carefully proven for the deformed dressing phase in appendix \ref{app:dressing phase}, which goes to zero in the relativistic limit. The only `coupling constant' left in the game is the level $k$, and as in the interpolating theory its effect comes in through the boundaries on the Y-system itself rather than the discontinuity relations the Y-system is supplemented with. It should not be surprising that the kernels in the simplified TBA equations do not depend explicitly on the coupling constant; this is a common feature of relativistic models. The information on the value of the coupling constant comes in through the kernels of the canonical TBA equations which do depend on it, which in turn define the asymptotics of the Y-functions.

There is one special case however, which is $Y_+$. Before taking the limit, $Y_+$ can be obtained as the analytic continuation of $Y_-$ through its cut along the real line. Now in the relativistic limit this cut disappears and $Y_+$ is a completely independent function that truly does not have a Y-system. It can be expressed in terms of $Y_-$ and $Y_Q$ as
\begin{equation}
Y_+(u) = Y_-(u) e^{\log\left(1+Y_Q \right) \star K_{Qy} (u)} \, ,
\end{equation}
where in the relativistic limit
\begin{equation}
K_{Qy} (u) = \frac{1}{2\pi i} \frac{d}{du} \log \frac{\sinh\frac{\pi}{4k}(u-i Q)}{\sinh\frac{\pi}{4k}(u+i Q)}\frac{\cosh\frac{\pi}{4k}(u+i Q)}{\cosh\frac{\pi}{4k}(u-i Q)}\, .
\end{equation}
In general the relation between $Y_+$ and $Y_-$ will have state dependent driving terms because of the convolution involving $Y_Q$. Applying $s^{-1}$ to the equation for $Y_+$ does not appear to give more insight.

Apart from this subtlety, the Y-system becomes a finite set of algebraic relations between meromorphic functions on the $u$-plane. Of course, also in this limit the Y-system depends on the state and twist in the manner described above.

\paragraph{Transfer matrix.} To obtain the tranfer matrix in the relativistic
limit we rescale the rapidities $u\rightarrow u/g,\ldots$. Because of this, the
factors $\SH^{\mathrm{II}}$ and $\SH^{\mathrm{III}}$ are basically unchanged
\begin{align}
\SH^{A}_B \rightarrow \prod_{i=1}^{K^A}\sinh\frac{\pi }{2k}(u-u_i ^{(A)} + B
i)\equiv \SSH^{A}_B.
\end{align}
Similarly, let us introduce the following functions for the $\KK{I}$ roots
\begin{align}
&\SSH^\mathrm{I}_B=\prod_{i=1}^{\KK{I}}\sinh\frac{\pi }{4k}(u-u_i+B i),
&&\tilde{\SSH}^\mathrm{I}_B = \prod_{i=1}^{\KK{I}}\cosh\frac{\pi }{4k}(u-u_i + B
i).
\end{align}
Then, we find that, after discarding the overall factor
$\prod_{i=1}^{K^{\mathrm{II}}} q^{\frac{a}{2}}\frac{y_i - x^-}{y_i -
x^+}\sqrt{\frac{x^+}{x^-}}$, the transfer matrix \eqref{fulltransfer} reduces to
\begin{align}
&t_{a,1}=
\sum_{m=0}^{a}\!
e^{i\alpha (a-2m)}
\frac{\SSH^\mathrm{I}_{a-2m+1}\tilde{\SSH}^\mathrm{I}_{a-2m-1}}{\SSH^\mathrm{I}_
{a+1}\tilde{\SSH}^\mathrm{I}_{a-1}}
\frac{\SSH^\mathrm{II}_a}{ \SSH^\mathrm{II}_{a-2m}}
+
\sum_{m=1}^{a-1}\!
e^{i\alpha(a-2m)}
\frac{\SSH^\mathrm{I}_{a-2m-1}\tilde{\SSH}^\mathrm{I}_{a-2m+1}}
{\SSH^\mathrm{I}_{a+1}\tilde{\SSH}^\mathrm{I}_{a-1}}
\frac{\SSH^\mathrm{II}_a}{ \SSH^\mathrm{II}_{a-2m}}
\\
&\quad -\! \sum_{m=0}^{a-1}\!
e^{i\alpha(a-2m-1)}
\frac{\SSH^\mathrm{I}_{a-2m-1}\tilde{\SSH}^\mathrm{I}_{a-2m-1}}
{\SSH^\mathrm{I}_{a+1}\tilde{\SSH}^\mathrm{I}_{a-1}}
\!\left[\!
e^{i\beta}
\frac{\SSH^\mathrm{II}_{a}\SSH^\mathrm{III}_{a-2m+1}}{\SSH^\mathrm{II}_{a-2m
}\SSH^\mathrm{II}_{a-2m-1} }
+e^{-i\beta}
\frac{\SSH^\mathrm{II}_{a}\SSH^\mathrm{III}_{a-2m-2}}{\SSH^\mathrm{II}_{a-2m
-2}\SSH^\mathrm{II}_{a-2m-1} }
\right]\!\nonumber.
\end{align}
Analogously, we derive the following expression for \eqref{eqn;Ts} in the
relativistic limit
\begin{align}
t_{1,s}  =&
\frac{\SSH^\mathrm{III}_{1-s}}{e^{i\beta s}\SSH^\mathrm{II}_{-s}}\left[
\sum_{m=0}^{s}
\frac{e^{2 i\beta m}\SSH^\mathrm{II}_{s-2m}\SSH^\mathrm{III}_{-1-s}}
{\SSH^\mathrm{III}_{s-2m-1}\SSH^\mathrm{III}_{s-2m+1}}
-e^{i\alpha}
\frac{\SSH^\mathrm{I}_{s+1}}{\SSH^\mathrm{I}_{s-1}}
\sum_{m=1}^{s}
\frac{e^{i\beta (2m-1)}
\SSH^\mathrm{II}_{s-2m}\SSH^\mathrm{III}_{-1-s}\SSH^\mathrm{III}_{s-1}}
{\SSH^\mathrm{III}_{s-2m-1}\SSH^\mathrm{III}_{s-2m+1}\SSH^\mathrm{III}_{s+1}}
+\right.
\\
&\left.
\frac{\SSH^\mathrm{I}_{s+1}\tilde{\SSH}^\mathrm{I}_{-s-1}}{\SSH^\mathrm{I}_{s-1}
\tilde{\SSH}^\mathrm{I }_{-s+1}}
\sum_{m=1}^{s-1}
\frac{e^{2i\beta m}
\SSH^\mathrm{II}_{s-2m}\SSH^\mathrm{III}_{1-s}\SSH^\mathrm{III}_{ s-1}}
{\SSH^\mathrm{III}_{s-2m-1}\SSH^\mathrm{III}_{s-2m+1}\SSH^\mathrm{III}_{s+1}}
-e^{-i\alpha}
\frac{\tilde{\SSH}^\mathrm{I}_{-s-1}}{\tilde{\SSH}^\mathrm{I}_{-s+1}}
\sum_{m=0}^{s-1}
\frac{e^{i\beta (2m+1)}\SSH^\mathrm{II}_{s-2m}\SSH^\mathrm{III}_{1-s}}
{\SSH^\mathrm{III}_{s-2m-1}\SSH^\mathrm{III}_{s-2m+1}}
\right].\nonumber
\end{align}
For completeness, let us also give the special functions $\tilde{t}$ used in
the asymptotic solution
\begin{align}
\tilde{t}_{a,1}=&
e^{i\alpha a}
\frac{\SSH^\mathrm{I}_{-a-1}\tilde{\SSH}^\mathrm{I}_{-a-3}}{\SSH^\mathrm
{I}_{a+1}\tilde{\SSH}^\mathrm{I}_{a-1}}
\frac{\SSH^\mathrm{II}_a}{ \SSH^\mathrm{II}_{-a-2}}
+
e^{i\alpha (a+2)}
\frac{\SSH^\mathrm{I}_{-a-1}\tilde{\SSH}^\mathrm{I}_{-a+1}}
{\SSH^\mathrm{I}_{a+1}\tilde{\SSH}^\mathrm{I}_{a-1}}
\frac{\SSH^\mathrm{II}_a}{ \SSH^\mathrm{II}_{-a}}
\\
& -
e^{i\alpha(a+1)}
\frac{\SSH^\mathrm{I}_{-a-1}\tilde{\SSH}^\mathrm{I}_{-a-1}}
{\SSH^\mathrm{I}_{a+1}\tilde{\SSH}^\mathrm{I}_{a-1}}
\!\left[\!
e^{i\beta}
\frac{\SSH^\mathrm{II}_{a}\SSH^\mathrm{III}_{-a+1}}{\SSH^\mathrm{II}_{-a}
\SSH^\mathrm{II}_{-a-1} }
+e^{-i\beta}
\frac{\SSH^\mathrm{II}_{a}\SSH^\mathrm{III}_{-a-2}}{\SSH^\mathrm{II}_{-a-2}
\SSH^\mathrm{II}_{-a-1} }
\right]\!\nonumber.\\
\tilde{t}_{1,s}  =&
e^{i\beta s}
\frac{\SSH^\mathrm{I}_{s+1}\tilde{\SSH}^\mathrm{I}_{-s-1}}{\SSH^\mathrm{I}
_{s-1}\tilde{\SSH}^\mathrm{I
}_{-s+1}}\frac{\SSH^\mathrm{III}_{s-1}\SSH^\mathrm{III}_{-s+1}}
{\SSH^\mathrm{III}_{-s-1}\SSH^\mathrm{III}_{s+1}}-
e^{i\alpha}e^{i\beta (s+1)}
\frac{\SSH^\mathrm{I}_{s+1}}{\SSH^\mathrm{I}_{s-1}}
\frac{
\SSH^\mathrm{II}_{-s-2}\SSH^\mathrm{III}_{s-1}\SSH^\mathrm{III}_{-s+1}}
{\SSH^\mathrm{II}_{-s}\SSH^\mathrm{III}_{-s-3}\SSH^\mathrm{III}_{s+1}}
+
\\
&
e^{i\beta (s+2)}
\frac{\SSH^\mathrm{I}_{s+1}\tilde{\SSH}^\mathrm{I}_{-s-1}}{\SSH^\mathrm{I}
_{s-1}\tilde{\SSH}^\mathrm{I }_{-s+1}}
\frac{
\SSH^\mathrm{II}_{-s-2}\SSH^\mathrm{III}_{-s+1}\SSH^\mathrm{III}_{
s-1}\SSH^\mathrm{III}_{-s+1}}
{\SSH^\mathrm{II}_{-s}\SSH^\mathrm{III}_{-s-3}\SSH^\mathrm{III}_{-s-1}
\SSH^\mathrm{III}_{s+1}}
-e^{-i\alpha}e^{i\beta (s+1)}
\frac{\tilde{\SSH}^\mathrm{I}_{-s-1}}{\tilde{\SSH}^\mathrm{I}_{-s+1}}
\frac{\SSH^\mathrm{III}_{-s+1}}
{\SSH^\mathrm{III}_{-s-1}}.\nonumber
\end{align}
Note that the transfer matrices are of purely trigonometric form.

\section{Concluding remarks}
\label{sec:conclusion}

In this paper we discussed the effects of excited states on the quantum deformed $\AdS$ mirror TBA equations. Doing so, we uncovered an interesting feature whereby the Y-system depends on (the excitation numbers of) the state under consideration, albeit in a mild way. This feature depends crucially on the root of unity deformation and the fact that we have a nested system, and to our knowledge has not been observed before. Similarly our TBA equations and Y-system depend explicitly on twisted boundary conditions, but this dependence is not unlike that in the XXZ spin chain, replacing twists by a magnetic field.

In the asymptotic limit these interesting results can be verified through our asymptotic solution given in terms of the transfer matrix of the $q$-deformed Hubbard chain and the $q$-deformed mirror bound state dressing phase, both of which we have explicitly constructed. This asymptotic solution also allows us to construct excited state TBA equations. It would be interesting to see to what extent the deformation can qualitatively affect the analytic properties of the Y-functions for an excited state in addition to the new effects which remain at the level of the Y-system as discussed above. As we saw in this paper for example, the ground state Y-functions of the twisted deformed model are nearly identical to those of the undeformed twisted model.

Our equations reduce to a particularly simple form in the relativistic limit, where they are conjectured to describe the Pohlmeyer reduced superstring theory. As these equations still have a coupling constant dependence in the form of $k$, this limit is nontrivial and would be interesting to study in its own right. At this point however, it would be instructive to first understand the relation between the $q$-deformed S-matrix and the (perturbative) S-matrix of the Pohlmeyer reduced superstring in more detail. Interestingly, there might be various subtleties in the relation between these S-matrices to which the resulting Bethe and subsequent TBA equations would be (rather) insensitive.

From the point of view of finite size integrability it would also be interesting to consider the case where the deformation parameter $q$ is taken to be real, where the theory shows some interesting features already mentioned in our previous paper. From the point of view of condensed matter physics, both types of deformations are interesting deformations of the Hubbard and related models. We can address the thermodynamics of these models in the spirit of e.g. \cite{Frolov:2011wg} with minor modifications of our equations.

\section*{Acknowledgements} We would like to thank Z. Bajnok, N. Beisert and S. Frolov for valuable
discussions, and S. Frolov for valuable comments on the paper. G.A. acknowledges
support by the Netherlands Organization for Scientific Research (NWO) under the
VICI grant 680-47-602. The work by M.L. is
supported by the SNSF under project number 200021-137616.
The work by G.A and S.T. is also part of the ERC Advanced grant research
programme No. 246974,  {\it ``Supersymmetry: a
window to non-perturbative physics"}.

\appendix

\section{Representations and S-matrix}

The asymptotic solution of the Y-system will be formulated in terms of transfer
matrices which partly take value in bound state representations. In this section
we will set some notation and discuss these representations together with the
corresponding S-matrix.

\subsection{Bound state representations}
\label{app:boundstatereps}

The quantum deformed S-matrix respects the quantum affine centrally extended
$\alg{psu}_q(2|2)$ algebra. This algebra is generated
by four sets of Chevalley generators $E_i$, $F_i$, $K_i(=q^{H_i})$ and two sets
of central elements $U_{k},V_{k}$ ($k=2,\,4$)
with $U_{k}$ being responsible for the braiding of the coproduct. The generators
1 and 3 form two quantum deformed $\alg{su}(2)$ algebras.

The symmetric matrix $DA$ and the normalization matrix $D$ associated to the
Cartan matrix $A$ are
\begin{equation}
DA=\begin{pmatrix}2 & -1 & 0 & -1\\
-1 & 0 & 1 & 0\\
0 & 1 & -2 & 1\\
-1 & 0 & 1 & 0
\end{pmatrix},\qquad D=\mathrm{diag}(1,-1,-1,-1).\label{DA}
\end{equation}
The algebra is then defined by the following non-trivial commutation relations,
\begin{align}
 & K_{i}E_{j}=q^{DA_{ij}}E_{j}K_{i}, &  & K_{i}F_{j}=q^{-DA_{ij}}F_{j}K_{i},\el
 & \{E_{2},F_{4}\}=-\tilde{g}\tilde{\alpha}^{-1}(K_{4}-U_{2}U_{4}^{-1}K_{2}^{-1}), &  & \{E_{4},F_{2}\}=\tilde{g}\tilde{\alpha}^{+1}(K_{2}-U_{4}U_{2}^{-1}K_{4}^{-1}),\el
 & [E_{j},F_{j}\}=D_{jj}\frac{K_{j}-K_{j}^{-1}}{q-q^{-1}}, &  & [E_{i},F_{j}\}=0,\quad i\neq j,\ i+j\neq6.
\end{align}
These are supplemented by Serre relations ($j=1,3,\, k=2,4$)
\begin{align}
 & [E_{j},[E_{j},E_{k}]]-(q-2+q^{-1})E_{j}E_{k}E_{j}=0, && [E_{1},E_{3}]=E_{2}E_{2}=E_{4}E_{4}=\{E_{2},E_{4}\}=0, \el
 & [F_{j},[F_{j},F_{k}]]-(q-2+q^{-1})F_{j}F_{k}F_{j}=0, && [F_{1},F_{3}]=F_{2}F_{2}=F_{4}F_{4}=\{F_{2},F_{4}\}=0.
\end{align}
The central elements are related to the quartic Serre relations (for $k=2,\,4$),
\begin{align}
 & \{[E_{1},E_{k}],[E_{3},E_{k}]\}-(q-2+q^{-1})E_{k}E_{1}E_{3}E_{k}=g_{k}\alpha_{k}(1-V_{k}^{2}U_{k}^{2}),\el
 & \{[F_{1},F_{k}],[F_{3},F_{k}]\}-(q-2+q^{-1})F_{k}F_{1}F_{3}F_{k}=g_{k}\alpha_{k}^{-1}(V_{k}^{-2}-U_{k}^{-2}).
\end{align}
In total, this algebra has three central charges,
\begin{eqnarray}
C_{1} & = & K_{1}K_{2}^{2}K_{3},\el
C_{2} & = & \{[E_{2},E_{1}],[E_{2},E_{3}]\}-(q-2+q^{-1})E_{2}E_{1}E_{3}E_{2},\el
C_{3} & = & \{[F_{2},F_{1}],[F_{2},F_{3}]\}-(q-2+q^{-1})F_{2}F_{1}F_{3}F_{2}.\label{C123}
\end{eqnarray}
Finally, the central elements $V_{k}$ are defined by the relation
$K_{1}^{-1}K_{k}^{-2}K_{3}^{-1}=V_{k}^{2}.$


\paragraph{Hopf algebra.}

The coproduct of the group-like elements $X\in\{1,K_{j},U_{k},V_{k}\}$
($j=1,2,3,4$ and $k=2,4$) is defined in the usual way, $\Delta(X)=X\otimes X$.
The remaining Chevalley generators have a coproduct  that is braided by the
central elements $U_{k}$
\begin{equation}
\Delta(E_{j})=E_{j}\otimes1+K_{j}^{-1}U_{2}^{+\delta_{j,2}}U_{4}^{+\delta_{j,4}}\otimes E_{j}\,,\quad\Delta(F_{j})=F_{j}\otimes K_{j}+U_{2}^{-\delta_{j,2}}U_{4}^{-\delta_{j,4}}\otimes F_{j}\,.\label{copEF}
\end{equation}


\paragraph{Representation.}

Following \cite{LMR} will use the $q$-oscillator representation to describe the
bound state representations (short symmetric represenatations). The bound-state
representation is defined on
vectors
\begin{align}
|m,n,k,l\rangle=(\mathsf{a}_{3}^{\dag})^{m}(\mathsf{a}_{4}^{\dag})^{n}(\mathsf{a}_{1}^{\dag})^{k}(\mathsf{a}_{2}^{\dag})^{l}\,|0\rangle,
\end{align}
where the indices $1$, $2$ denote bosonic and $3$, $4$ - fermionic oscillators.
The total number of excitations $k+l+m+n=M$ is the bound-state number and the
dimension of the representation is dim$\,=\!4M$.  This representation constrains
the central elements via $U:=U_2=U^{-1}_4$  and $V:=V_2=V^{-1}_4$ and describes
an excitation with quasi-momentum $p$ related to the deformation
parameter as $U^2=e^{ip}$.

The explicit action of the triples corresponding to the bosonic and fermionic $\mathfrak{sl}_{q}(2)$ in this representation are given by
\begin{align}
 & H_{1}|m,n,k,l\rangle=(l-k)|m,n,k,l\rangle, &  & H_{3}|m,n,k,l\rangle=(n-m)|m,n,k,l\rangle,\el
 & E_{1}|m,n,k,l\rangle=[k]_{q}\,|m,n,k-1,l+1\rangle, &  & E_{3}|m,n,k,l\rangle=|m+1,n-1,k,l\rangle,\el
 & F_{1}|m,n,k,l\rangle=[l]_{q}\,|m,n,k+1,l-1\rangle, &  &
F_{3}|m,n,k,l\rangle=|m-1,n+1,k,l\rangle.
\end{align}
The supercharges act on basis states as
\begin{align}
H_{2}|m,n,k,l\rangle= & ~-\left\{ C-\frac{k-l+m-n}{2}\right\} |m,n,k,l\rangle,\el
E_{2}|m,n,k,l\rangle= & ~a~(-1)^{m}[l]_{q}\,|m,n+1,k,l-1\rangle+b~|m-1,n,k+1,l\rangle,\el
F_{2}|m,n,k,l\rangle= & ~c~[k]_{q}\,|m+1,n,k-1,l\rangle+d~(-1)^{m}\,|m,n-1,k,l+1\rangle.
\end{align}
Here $[n]_q=(q^n-q^{-n})/(q-q^{-1})$ denotes the $q$-number and $C$ is the
$q$-factor of the central element $V = q^C$ and represents the energy of the
state.

The explicit parametrization of the representation labels in terms of the
conventional $x^{\pm}$  variables is
\begin{align}
a & =\sqrt{\frac{g}{[M]_{q}}}\gamma\,, && b=\sqrt{\frac{g}{[M]_{q}}}\frac{\alpha}{\gamma}\frac{x^{-}-x^{+}}{x^{-}}\,,\el
c & =\sqrt{\frac{g}{[M]_{q}}}\frac{\gamma}{\alpha\, V}\frac{i\,\tilde{g}\, q^{\frac{M}{2}}}{g(x^{+}+\xi)}\,, &  & d=\sqrt{\frac{g}{[M]_{q}}}\frac{\tilde{g}\, q^{\frac{M}{2}}V}{i\, g\,\gamma}\frac{x^{+}-x^{-}}{\xi x^{+}+1}\,.\label{abcd}
\end{align}
The central elements in this parametrization read as
\begin{align}
&U^2 = \frac{1}{q^M} \frac{x^+ + \xi}{x^- + \xi} =  q^M \frac{x^+}{x^-}\frac{\xi x^- + 1}{\xi x^+ + 1},
  && V^2 = \frac{1}{q^M} \frac{\xi x^+ + 1}{\xi x^- + 1} = q^M \frac{x^+}{x^-}\frac{x^- + \xi}{x^+ + \xi},
\end{align}
while the shortening condition becomes
\begin{align}
 \frac{1}{q^{M}}\left(x^+ + \frac{1}{x^+} + \xi + \frac{1}{\xi}\right) = q^M\left(x^- + \frac{1}{x^-}+ \xi + \frac{1}{\xi}\right) ,
\end{align}
here $\xi = -i \tilde g (q-q^{-1})$ and $\tilde g^2={g^2}/({1-g^2(q-q^{-1})^2})$.

The action of the affine charges $H_{4}$, $E_{4}$, $F_{4}$ is defined exactly the same as the regular supercharges, but is subject to the following substitutions $C\to-C$ and $(a,b,c,d)\to(\tilde{a},\tilde{b},\tilde{c},\tilde{d}).$  The affine labels $\tilde{a},\tilde{b},\tilde{c},\tilde{d}$ are acquired from (\ref{abcd}) by replacing
\begin{align}
V\rightarrow V^{-1}, \qquad
x^{\pm}\rightarrow\frac{1}{x^{\pm}}, \qquad
\gamma\rightarrow\frac{i\tilde{\alpha}\gamma}{x^{+}}, \qquad
\alpha\rightarrow\alpha\,\tilde{\alpha}^{2}.
\end{align}
Let us introduce the multiplicative spectral (evaluation) parameter $\U$ and
a multiplicative parameter $u$ of the algebra
\begin{align}\label{z}
&\U = \frac{1-U^2V^2}{V^2-U^2}  = e^{\frac{g\pi u}{k}}.
\end{align}
To keep the the expressions in the remainder of the paper manageable, we introduce the following short-hand notation for the basis vectors of the representation
\begin{align}\label{eqn;basise}
&|e_k\rangle = |0,0,M-k,k\rangle, &&|e_{3,k}\rangle = |0,1,M-k-1,k\rangle, \nonumber\\
&|e_{4,k}\rangle = |1,0,M-k-1,k\rangle, &&|e_{34,k}\rangle = |1,1,M-k-1,k-1\rangle
\end{align}
and we suppress the total bound state number $M$.

\subsection{S-matrix}

We consider the bound state S-matrix which is an intertwining matrix of
the tensor space spanned by the vectors
\begin{align}\label{bstates}
|m_1,n_1,k_1,l_1\rangle\otimes|m_2,n_2,k_2,l_2\rangle.
\end{align}
Here $0 \leq m_1, n_1, m_2, n_2 \leq 1$ and $k_1, l_1, k_2, l_2 \geq 0$ denote
the numbers of fermionic and bosonic excitations respectively with the bound
state number $M_i$ being the total number of excitations $M_i = m_i + n_i + k_i
+ l_i$. The S-matrix is required to be invariant under the coproducts of the
affine algebra,
\begin{align}\label{Sinv}
\S\,\Delta(J) = \Delta^{op}(J)\, \S, \qquad\text{for any}\qquad J\in\afQ.
\end{align}
The invariance under the bosonic symmetries $\DH_1$ and $\DH_3$ requires
the total number of fermions and the total number of fermions of one
type\footnote{Note that a bosonic excitation is interpreted as a combined
excitation of two fermions of different type.}
\begin{align}
N_f &= m_1 + m_2 + n_1 + n_2 + 2 l_1 + 2 l_2,\el
N_{f_3} &= m_1 + m_2 + l_1 + l_2.\label{Nf}
\end{align}
to be conserved. This conservation divides the space \eqref{bstates} into
invariant subspaces denoted I,II,III in \cite{LMR}. The explicit
expressions for the coefficients are given in \cite{LMR}. In this paper we will
restrict to the case where $M_2=1$ and we have for completeness included the
S-matrix elements that are needed in our derivations in Appendix
\ref{app;Smatrix}

\section{Algebraic Bethe Ansatz}

In this appendix we give various details concerning the algebraic Bethe Ansatz
preformed in Section \ref{sec;ABA}.

\subsection{Elements of the S-matrix}\label{app;Smatrix}

In order to derive the transfer matrix we need the following elements of the
S-matrix where the first leg is an $M$-particle bound state and the second leg
is in a fundamental representation. The S-matrix that scatters subspace I is
\begin{align}
&\mathscr{X}^{m,0}_{m} = q^{\frac{M-1}{2}}  \frac{x^-_1 - x^+_2}{x^+_1 - x^-_2}  \frac{U_1V_1}{U_2V_2}.
\end{align}
For subspace II we have
\begin{align}
&\mathscr{Y}^{m,0;1}_{m;1} = \frac{q^{m-\frac{1}{2}}}{U_2V_2} \frac{x^+_1 - x^+_2}{x^+_1 - x^-_2}  \frac{q^{M-2m} \U_1 - q \U_2}{q^M \U_1 - q \U_2}
\\
&\mathscr{Y}^{m,0;1}_{m;2} = q^m \frac{[M-m]_q}{\sqrt{[M]_q}} \frac{\gamma_1}{\gamma_2} \frac{x^+_2-x^-_2}{x^+_1 -x^-_2}\\
&\mathscr{Y}^{m,0;2}_{m;1} = \frac{q^{\frac{M-1}{2}-m}}{[M]_q}\frac{\gamma_2}{\gamma_1} \frac{x^+_1-x^-_1}{x^+_1 -x^-_2}\sqrt{\frac{x^+_1x^-_2}{x^-_1x^+_2}}\\
&\mathscr{Y}^{m,0;2}_{m;2} = q^{\frac{M}{2}} \frac{x^-_1-x^-_2}{x^+_1 -x^-_2}\sqrt{\frac{x^+_1}{x^-_1}}\\
&\mathscr{Y}^{m,0;1}_{m;4} = \frac{q^{m}\alpha}{\gamma_1\gamma_2\sqrt{[M]_q}}
\frac{(x^+_1-x^-_1)(x^+_2-x^-_2)(x^-_1-x^+_2)x^+_1}{(x_1^+x^+_2-1)(x^+_1 -x^-_2) x_1^-}\\
&\mathscr{Y}^{m,0;4}_{m;1} = \frac{[m]_q}{\sqrt{[M]_q}}\frac{\gamma_1\gamma_2}{\alpha}\frac{q^{\frac{3M-1}{2}-m}}{x_1^+x^+_2-1}
\frac{x^-_1-x^+_2}{x^+_1-x^-_2}\sqrt{\frac{x^+_1x^-_2}{x^-_1x^+_2}}\\
&\mathscr{Y}^{m,0;4}_{m;4} =q^{\frac{M}{2}} \frac{1-x^+_1x^-_2}{1-x^+_1x^+_2} \frac{x^-_1-x^+_2}{x^+_1-x^-_2} \sqrt{\frac{x^+_1}{x^-_1}}.
\end{align}
Finally
\begin{align}
&\mathscr{Z}^{m,0;1}_{m;1} = q^m\left[1 - q^{M-m}\frac{[m]_q}{[M]_q} \frac{(x^+_1-x^-_1)(x^-_1 x^+_1 x^+_2 -x^-_2)}{(x^+_1-x^-_2)(x^-_1 x^+_1 x^+_2 -x^-_1)}\right]\\
&\mathscr{Z}^{m,0;3}_{m;1} = \frac{[m]_q[M-m]_q}{[M]_q} \frac{x^+_2-x^-_2}{x^+_1-x^-_2}\frac{q^M}{1-x^+_1x^+_2}\frac{\gamma_1^2}{\alpha}\\
&\mathscr{Z}^{m,0;1}_{m;3} = \frac{1}{[M]_q}\frac{(x^+_1 - x^-_1)^2 (x^+_2-x^-_2)x^+_1}{(x^+_1 - x^-_2) (1 - x^+_1x^+_2)x^-_1} \frac{\alpha}{\gamma_1^2}\\
&\mathscr{Z}^{m,0;3}_{m;3} = q^m\left[\frac{(x^+_1-x^+_2)(1-x^+_1x^-_2)}{(x^+_1-x^-_2)(1-x^+_1x^+_2)} -
q^{M-m}\frac{[m]_q}{[M]_q} \frac{(x^+_1-x^-_1)(x^-_1 x^+_1 x^-_2 -x^-_2)}{(x^+_1-x^-_2)(x^-_1 x^+_1 x^+_2 -x^-_1)} \right]\\
&\mathscr{Z}^{m,0;6}_{m;6} = q^{\frac{M+1}{2}}\frac{ x^-_1-x^-_2}{x^+_1 -x^-_2} \frac{ 1- x^+_1 x^-_2}{1- x^+_1x^+_2}  \sqrt{\frac{x^+_1x^+_2}{x^-_1x^-_2}}\\
&\mathscr{Z}^{m,0;3}_{m;6} = q^{m+\frac{1}{2}} \frac{\gamma_1}{\gamma_2} \frac{[M-m]_q}{\sqrt{[M]_q}} \frac{ x^+_2-x^-_2}{x^+_1 -x^-_2} \frac{ 1- x^+_1 x^-_2}{1- x^+_1x^+_2}  \sqrt{\frac{x^+_2}{x^-_2}}\\
&\mathscr{Z}^{m,0;1}_{m;6} = \frac{q^{m+\frac{1}{2}}}{\sqrt{[M]_q}} \frac{\alpha}{\gamma_1\gamma_2}
\frac{(x^-_1-x^-_2) (x^+_1-x^-_1)(x^+_2-x^-_2)}{(x^+_1 -x^-_2)(1-x^+_1x^+_2)}\frac{x^+_1}{x^-_1}\sqrt{\frac{x^+_2}{x^-_2}}\\
&\mathscr{Z}^{m,0;6}_{m;1} = \frac{q^{\frac{M}{2}-m}}{\sqrt{[M]_q}} \frac{\gamma_2} {\gamma_1}
\frac{x^-_1-x^-_2}{(x^+_1 -x^-_2)(1-x^+_1x^+_2)}\sqrt{\frac{x^+_1}{x^-_1}}\\
&\mathscr{Z}^{m,0;6}_{m;3} = \frac{q^{\frac{3M}{2}-m}[m]_q}{\sqrt{[M]_q}} \frac{\gamma_1\gamma_2} {\alpha}
\frac{x^+_1-x^-_1}{x^+_1 -x^-_2}\frac{1-x^+_1 x^-_2}{1-x^+_1x^+_2}\sqrt{\frac{x^+_1}{x^-_1}}.
\end{align}

\subsection{Commutation relations}\label{app;commrel}

Here we list the needed commutation relations. Greek indices run over the
fermionic indices 3,4. The diagonal elements $T^I_I$ depend on the spectral
parameter $z$ and the representation is described by the parameters $x^\pm$
without indices. The creation operator depends on the spectral parameter
$\lambda$ and we further denote $x^+(\lambda)=y$ and $\V = q \lambda$
\begin{align}
T^0_0 T^1_\alpha(\lambda) &=   \left[q^{\frac{a}{2}} \frac{y-x^-}{y-x^+}
\sqrt{\frac{x^+}{x^-}} \right]\! T^1_\alpha(\lambda) T^0_0 +\ldots\\
T^a_a T^1_\alpha(\lambda) &=   \left[q^{\frac{a}{2}} \frac{y-x^-}{y-x^+} \sqrt{\frac{x^+}{x^-}} \right]\! \frac{q^a \U - \V}{\U - q^a \V}T^1_\alpha(\lambda) T^a_a+\ldots\\
(T^m_m + T^{34,m}_{34,m}) T^1_\alpha(\lambda) &= \left[q^{\frac{a}{2}} \frac{y-x^-}{y-x^+} \sqrt{\frac{x^+}{x^-}} \right]\! \frac{q^a \U - \V}{q^{a-m} \U - q^m \V} T^1_\alpha(\lambda) (T^m_m + T^{34,m}_{34,m})+\ldots \\
T^{\beta,m}_{\beta,m} T^1_\alpha(\lambda) &= \left[q^{\frac{a}{2}} \frac{y-x^-}{y-x^+} \sqrt{\frac{x^+}{x^-}} \right]\! \frac{q^a \U - \V}{q^{a-m} \U - q^m \V} r_{\beta\alpha}^{\gamma\delta}(q^{M-2m-1}\U,\lambda) T^1_\delta(\lambda)T^{\beta,m}_{\gamma,m},
\end{align}
where the dots stand for additional terms that do not affect the eigenvalue but should vanish upon solving the auxiliary Bethe equations. The auxiliary S-matrix $r$ is given by \eqref{eqn;auxr}.

\subsection{Auxiliary Bethe Ansatz}\label{app;auxBAE}

Consider a general $\alg{gl}(1)^2$ invariant S-matrix\begin{align}
r(x,y) =
\begin{pmatrix}
r_1(x,y) & 0 & 0 & 0 \\
0 & r_2(x,y) & r_5(x,y) & 0 \\
0 & r_6(x,y) & r_3(x,y) & 0 \\
0 & 0 & 0 & r_4(x,y)
\end{pmatrix},
\end{align}
satisfying the Yang-Baxter equation and unitarity (\textit{i.e.}
$r_{12}(x_1,x_2)r_{21}(x_2,x_1)=1$). Following the algebraic Bethe ansatz
approach we introduce the monodromy ($M$), transfer matrix ($T$) and vacuum
state ($|0\rangle$) on a chain with $L$ sites
\begin{align}
&M = \begin{pmatrix} A & B \\ C & D \end{pmatrix} && T = \mathrm{tr} M = A+D, && |0\rangle = \bigotimes_{i=1}^{L}\begin{pmatrix}1\\0 \end{pmatrix}
\end{align}
The action of of the different elements of the monodromy matrix on the vacuum
are readily computed
\begin{align}
& A(q)|0\rangle = \prod_{i=1}^L r_1(q,p_i)|0\rangle, && D(q)|0\rangle = \prod_{i=1}^L r_3(q,p_i)|0\rangle, && C|0\rangle = 0.
\end{align}
And the operators from the monodromy matrix are found to satisfy the following
commutation relations
\begin{align}
&A(x)B(y) = \frac{r_1(y,x)}{r_3(y,x)} B(y)A(x) - \frac{r_5(y,x)}{r_3(y,x)} B(x)A(y), \\
&D(x)B(y) = \frac{r_4(x,y)}{r_3(x,y)} D(y)B(x) - \frac{r_6(x,y)}{r_3(x,y)} D(x)B(y), \\
& B(x)B(y) = \frac{r_1(x,y)}{r_4(x,y)} B(y)B(x).
\end{align}
For a state with $K$ excitations we make the ansatz
\begin{align}
B(\lambda_1)\ldots B(\lambda_K)|0\rangle.
\end{align}
Then putting all the ingredients together we find the following eigenvalue for
the transfer matrix
\begin{align}
\Lambda(q|\vec{p},\vec{\lambda}) = \prod_{i=1}^L r_1(q,p_i) \prod_{i=1}^K  \frac{r_1(\lambda_i,q)}{r_3(\lambda_i,q)}  +  \prod_{i=1}^L r_3(q,p_i) \prod_{i=1}^K  \frac{r_4(q,\lambda_i)}{r_3(q,\lambda_i)},
\end{align}
supplemented with the Bethe equations
\begin{align}
1=-\frac{r_5(\lambda_k,q) r_3(q,\lambda_k)}{r_3(\lambda_k,q) r_6(q,\lambda_k)} = \prod_{i=1}^L \frac{r_1(\lambda_k,p_i)}{r_3(\lambda_k,p_i)}
\prod_{i\neq k}^K \frac{r_1(\lambda_i,\lambda_k)r_3(\lambda_k,\lambda_i)}{r_3(\lambda_i,\lambda_k)r_4(\lambda_k,\lambda_i)},
\end{align}
where we point out that the LHS is actually constant due to unitarity.

\section{Dualization}\label{app;dual}

There is a way to relate the transfer matrices and Bethe equations in different
gradings that goes under the name of dualization. They are related to each other
by a change in variables and quantum numbers. Here we will apply a duality
transformation to the $y$ roots, giving transfer matrix in the $\alg{su}(2)$
and $\alg{sl}(2)$ gradings.

The transfer matrix in the $\alg{sl}(2)$ is given by \eqref{fulltransfer}. Let
us introduce the polynomial of degree $K^{\mathrm{I}} + 2
K^{\mathrm{III}}$ in terms of the multiplicative evaluation parameter
\begin{align}
P(y) = \prod_{i=1}^{K^\mathrm{I}}\! \frac{y-x^+_i}{\sqrt{x^+_i}} \prod_{i=1}^{K^\mathrm{III}}\! y (\W_i - q \V) -
q^{\frac{K^{\mathrm{I}}}{2}}\prod_{i=1}^{K^\mathrm{I}}\! \frac{y-x^-_i}{\sqrt{x^-_i}} \prod_{i=1}^{K^\mathrm{III}}\! y (q \W_i - \V) \equiv P_1(y) - P_2(y),
\end{align}
where $\V = \frac{y+y^{-1} + \xi + \xi^{-1}}{\xi^{-1}-\xi}$. By construction the
polynomial $P$ has $K^{\mathrm{II}}$ roots corresponding to the roots of the
auxiliary Bethe equations \eqref{eq:auxyfinal}. Factorizing $P$ then yields
\begin{align}
P(y)  = c  \prod_{i=1}^{K^\mathrm{II}}  (y-y_i) \prod_{i=1}^{\tilde{K}^\mathrm{II}}  (y-\tilde{y}_i),
\end{align}
where $\tilde{y}_i$ are the dual roots and there are $\tilde{K}^\mathrm{II}$
of them. As a consequence we find the following identity
\begin{align}\label{eqn;dualroots}
 \prod_{i=1}^{K^\mathrm{II}}  \frac{y_i-a}{y_i-b} =  \frac{P(a)}{P(b)} \prod_{i=1}^{\tilde{K}^\mathrm{II}}  \frac{\tilde{y}_i-b}{\tilde{y}_i-a}.
\end{align}
Dualizing means to write the transfer matrix and Bethe eqautions in terms of
$\tilde{y}$ and $\tilde{K}^\mathrm{II}$. For conciseness we will present he
deriavtion in terms of the multiplicative parameters.

\paragraph{Auxiliary Bethe equations.} Let us now dualize the Bethe equations.
Since $\tilde{y}$ is a root of $P$ it satisfies the same auxiliary Bethe
equations \eqref{eq:auxyfinal}
\begin{align}
1&= \prod_{i=1}^{K^{\mathrm{I}}}\frac{1}{\sqrt{q}}\frac{\tilde{y}_k - x^+_i}{\tilde{y}_k - x^-_i}\sqrt{\frac{x^-_i}{x^+_i}}
\prod_{i=1}^{K^{\mathrm{III}}}\frac{\W_i - q \tilde{\V}_k}{q \W_i - \tilde{\V}_k}.
\end{align}
Next, we continue by looking at the dualization of \eqref{eq:auxwfinal}.
Introducing two auxiliary parameters $\omega^\pm$
\begin{align}
\W_k = - \frac{1}{q}  \frac{\omega^+ + \frac{1}{\omega^+} + \xi + \frac{1}{\xi}}{\xi-\xi^{-1}}
= - q  \frac{\omega^- + \frac{1}{\omega^-} + \xi + \frac{1}{\xi}}{\xi-\xi^{-1}},
\end{align}
and using \eqref{eqn;dualroots} then allows us to write
\begin{align}
\prod_{i=1}^{K^{\mathrm{II}}} \frac{q \W_k - \V_i}{\W_k - q \V_i}  =
\prod_{i=1}^{K^{\mathrm{II}}} \frac{1}{q} \frac{(y_i - \omega^+_k )(y_i - \frac{1}{\omega^+_k})}{(y_i - \omega^-_k )(y_i - \frac{1}{\omega^-_k})} =
\frac{1}{q^{K^\mathrm{II}}}\frac{P(\omega^+_k)P(\frac{1}{\omega^+_k})}{P(\omega^-_k)P(\frac{1}{\omega^+_k})}\prod_{i=1}^{\tilde{K}^{\mathrm{II}}} \frac{1}{q}\frac{\W_k - q \tilde{\V}_i} {q \W_k - \tilde{\V}_i}.
\end{align}
From the definition of $P$ it is straightforward to prove
\begin{align}
\frac{P(\omega^+_k)P(\frac{1}{\omega^+_k})}{P(\omega^-_k)P(\frac{1}{\omega^+_k})}  =
q^{K^\mathrm{I} + 2 K^\mathrm{III}}  \prod_{i=1}^{K^{\mathrm{III}}} \left(\frac{\W_i-q^2 \W_k}{q^2 \W_i - \W_k}\right)^2,
\end{align}
resulting in the following dualized version of  \eqref{eq:auxwfinal}
\begin{align}
-1 = \prod_{i=1}^{\tilde{K}^{\mathrm{II}}} \frac{\W_k - q \tilde{\V}_i} {q \W_k - \tilde{\V}_i} \prod_{i=1}^{K^{\mathrm{III}}} \frac{q^2 \W_k - \W_i}{\W_k - q^2 \W_i},
\end{align}
where we used that $\tilde{K}^\mathrm{II} = K^\mathrm{I} - K^\mathrm{II} +2
K^\mathrm{III}$. From this we see that the dualized auxiliary equations can be
simply obtained from \eqref{eq:auxyfinal} and \eqref{eq:auxwfinal} by replacing
$y\leftrightarrow \tilde{y} $ and $K^\mathrm{II} \leftrightarrow
\tilde{K}^\mathrm{II} $ respectively.

\paragraph{Transfer matrix.} After discarding the overall scalar factor and
rearranging some terms in the transfer matrix, we find
\begin{align}
&t_{a,1}=- \frac{P(x^+)}{P_2(x^+)} + \left(
\frac{R^{+}_{-a}B^{-}_{-a}}{R^{+}_a B^{-}_a}
\prod_{i=1}^{K^{\mathrm{II}}}\frac{q^a \U-\V_i}
{\U - q^a \V_i}\right)\frac{P(\frac{1}{x^-})}{P_1(\frac{1}{x^-})} + \\
&\frac{R^{-}_a}{R^{+}_a} \! \sum_{m=1}^{a-1} \!
\frac{P(x^{[a-2m]})P(\frac{1}{x^{[a-2m]}})}{P_1(x^{[a-2m]})P_1(\frac{1}{x^{[a-2m
]}})} \prod_{i=1}^{K^{\mathrm{I}}}\! \frac{q^{a-m} \U - q^{m+1} \U_i }{q^{a} \U
- q \U_i} \prod_{i=1}^{K^{\mathrm{II}}}\frac{q^{a} \U-\V_i}{q^{a-k}\U - q^{m}
\V_i} \prod_{i=1}^{K^{\mathrm{III}}}\! \frac{\U q^{a-2m+1}-\W_i }{\U
q^{a-2m}-\W_i q}.
\nonumber
\end{align}
We can write this in a more convenient form in terms of the dual quantum numbers
by using the following properties of $P$
\begin{align}
&\prod_{i=1}^{K^{\mathrm{I}}}\frac{q^{a-2m}\U - q \U_i }{q^{a} \U - q \U_i} =
\frac{ P_1(x^{[a-2m]})P_1(\frac{1}{x^{[a-2m]}})}{P_1(x^+)P_1(\frac{1}{x^+})}
\prod_{i=1}^{K^{\mathrm{III}}} \left(\frac{q^{a+1} \U - \W_i}{q^{a-2m+1} \U - \W_i}\right)^2,\\
&\prod_{i=1}^{K^{\mathrm{II}}}\frac{q^{a} \U-\V_i}{q^{a-m}\U - q^{m} \V_i}  =\frac{1}{q^{m (K^\mathrm{I} +2 K^\mathrm{III} - \tilde{K}^\mathrm{II}  )}} \frac{P(x^+)P(\frac{1}{x^+})}{P(x^{[a-2m]})P(\frac{1}{x^{[a-2m]}})}
\prod_{i=1}^{\tilde{K}^{\mathrm{II}}} \frac{q^{a-2m}\U - \tilde{\V}_i}{q^{a} \U - \tilde{\V}_i},
\end{align}
leading finally to
\begin{align}
&\hat{t}_{a,1}= \frac{P(x^+)}{P_2(x^+)}  \frac{P_2(x^-)}{P(x^-)}  \left[- \frac{P(x^-)}{P_2(x^-)} +
\frac{P(\frac{1}{x^+})}{P_1(\frac{1}{x^+})}
\prod_{i=1}^{\tilde{K}^{\mathrm{II}}} \frac{\U - q^a \tilde{\V}_i}{q^{a}\U - \tilde{\V}_i}
\prod_{i=1}^{K^{\mathrm{III}}} \frac{(q^{a-1} \U - \W_i)(q^{a+1} \U - \W_i)}{(\U - q^{a-1} \W_i )(\U - q^{a+1} \W_i)}  + \right. \\
&\qquad+\left. \frac{P(\frac{1}{x^+})}{P_1(\frac{1}{x^{+}})} \frac{P(x^-)}{P_2(x^-)}  \sum_{m=1}^{a-1}
\prod_{i=1}^{\tilde{K}^{\mathrm{II}}} \frac{q^{a-m}\U - q^m\tilde{\V}_i}{q^{a} \U - \tilde{\V}_i}
\prod_{i=1}^{K^{\mathrm{III}}} \frac{(q^{a+1} \U - \W_i)(q^{a-1} \U - \W_i)}{( q^{a-m} \U - q^{m+1} \W_i)(q^{a-m} \U - q^{m-1} \W_i)}\right]\!.
\nonumber
\end{align}
This is the dualized transfer matrix.

\paragraph{Relating $\hat{t}_{a,1}$ to $t_{1,s}$.} Let us introduce the polynomial $\bar{P}$, which is the complex conjugate of $P$
\begin{align}
\bar{P}(y) = \prod_{i=1}^{K^\mathrm{I}}\! \frac{y-x^-_i}{\sqrt{x^-_i}} \prod_{i=1}^{K^\mathrm{III}}\! y (\W_i - \frac{\V}{q}) -
\frac{1}{q^{\frac{K^{\mathrm{I}}}{2}}}\prod_{i=1}^{K^\mathrm{I}}\! \frac{y-x^+_i}{\sqrt{x^+_i}} \prod_{i=1}^{K^\mathrm{III}}\! y ( \frac{\W_i}{q} - \V)
\equiv \bar{P}_1(y) - \bar{P}_2(y).
\end{align}
By similar arguments it can be shown that $T_{1,s}$ can be written as
\begin{align}
t_{1,s}  =& -\frac{\bar{P}(x^+)}{\bar{P}_2(x^+)}
+ \frac{\bar{P}(\frac{1}{x^-})}{\bar{P}_1(\frac{1}{x^-})} \prod_{i=1}^{K^{\mathrm{II}}} \frac{q^{s} \U - \V_i}{\U - q^s \V_i}  \prod_{i=1}^{K^{\mathrm{III}}} \frac{\U - q^{s+1}\W_i}{q^{s}\U - q \W_i}  \, \frac{q\U - q^{s}\W_i}{q^{s+1}\U - \W_i} + \\
& + \frac{\bar{P}(\frac{1}{x^-})\bar{P}(x^+)}{\bar{P}_1(\frac{1}{x^-})\bar{P}_2(x^+)}\sum_{m=1}^{s-1}
\prod_{i=1}^{K^{\mathrm{II}}} \frac{q^{s-m} \U - q^m \V_i}{\U - q^s \V_i}
\prod_{i=1}^{K^{\mathrm{III}}} \frac{(\U - q^{s+1}\W_i)(\U - q^{s-1}\W_i)}{(q^{s-m}\U - q^{m+1}\W_i)(q^{s-m}\U - q^{m-1}\W_i)} \nonumber
\end{align}
It is now trivial to see that $t_{1,s}$ maps into $\hat{t}_{a,1}$ (after changing $k\leftrightarrow s-k$) under complex conjugation (\textit{i.e.} map $(x^\pm,q) \rightarrow (x^\mp,1/q) $) up to an overall normalization.

\section{The dressing phase}

\label{app:dressing phase}

\subsection{The dressing phase for fundamental particles of the 'string' theory}

The crossing equation that follows from the $q$-deformed $R$-matrix has a solution which is a natural deformation of the $\ads$ dressing phase \cite{Arutyunov:2004vx,Beisert:2006ez,Dorey:2007xn}. This solution, which we will denote $\tilde{\sigma}$, was found in \cite{Hoare:2011wr}. The dressing phase in our conventions, $\sigma$, is related to $\tilde{\sigma}$ as
\begin{equation}
\label{eq:dressingvsdressing}
\sigma^2(x_1,x_2)  = \tilde{\sigma}^2 (x_1,x_2) \frac{x_1^+}{x_1^-} \frac{x_2^-}{x_2^+} \frac{x_1^- +\xi}{x_1^+ +\xi}\frac{x_2^+ +\xi}{x_2^- +\xi} \equiv  \tilde{\sigma}^2 (x_1,x_2) \frac{P(x_1)}{P(x_2)}\, .
\end{equation}
Both $\sigma$ and $\tilde{\sigma}$ solve the $\ads$ crossing equation \cite{Janik:2006dc} in the limit $q\rightarrow1$. The dressing phase $\tilde{\sigma}$ is conventionally written in the form
\begin{equation}
\tilde{\sigma}(z_1,z_1) \equiv e^{i \tilde{\theta}(z_1,z_1)} = \exp { i\left(\chi(x_1^+,x_2^+) - \chi(x_1^-,x_2^+) - \chi(x_1^+,x_2^-) + \chi(x_1^-,x_2^-)\right) } \, ,
\end{equation}
where when both particles are in the string region, the $\chi$ functions are given by
\begin{equation}
\chi(x_1,x_2) = i \oint_{|z|=1} \frac{dz}{2 \pi i} \frac{1}{z-x_1}\oint_{|w|=1}  \frac{1}{w-x_2}\frac{dw}{2 \pi i} \log  \frac{\Gamma_{q^2} (1+\frac{ig}{2}(u(z)-u(w)))}{\Gamma_{q^2} (1-\frac{ig}{2}(u(z)-u(w)))}\,.
\end{equation}
Here $\Gamma_q$ is the $q$-analogue of the $\Gamma$ function, which satisfies
\begin{equation}
\label{eq:Gammaqdef}
\Gamma_{q^2}(1+x) = \frac{1-q^{2x}}{1-q^2} \Gamma_{q^2}(x)\, .
\end{equation}
Under analytic continuation to other regions of the rapidity torus the expression for the $\chi$ function changes. The above double integral is commonly denoted $\Phi$
\begin{equation}
\label{eq:phidef}
\Phi(x_1,x_2) \equiv i \oint_{|z|=1} \frac{dz}{2 \pi i} \frac{1}{z-x_1}\oint_{|w|=1}  \frac{1}{w-x_2}\frac{dw}{2 \pi i} \log  \frac{\Gamma_{q^2} (1+\frac{ig}{2}(u(z)-u(w)))}{\Gamma_{q^2} (1-\frac{ig}{2}(u(z)-u(w)))}\, ,
\end{equation}
and is equal to the $\chi$ function in the string region. The $\Phi$ function has a discontinuity on the edge of the string region ($|x^\pm|=1$), and to properly define the $\chi$ function beyond it we will need further terms, which leads us to introduce the $\Psi$ function
\begin{equation}
\label{eq:psidef}
\Psi(x_1,x_2) \equiv i \oint_{|z|=1} \frac{dz}{2 \pi i} \frac{1}{z-x_2} \log  \frac{\Gamma_{q^2} (1+\frac{ig}{2}(u_1-u(z)))}{\Gamma_{q^2} (1-\frac{ig}{2}(u_1-u(z)))} \, .
\end{equation}
With these definitions we are ready to give the expression for the dressing phase in the currently relevant regions of the torus, defined as
\begin{equation}
\mathcal{R}_{0}: |x^\pm|>1\, , \, \, \, \mathcal{R}_{1}: |x^+|<1,|x^-|>1\, , \, \, \,  \mathcal{R}_{2}: |x^+|<1,|x^-|<1\, .
\end{equation}
$\mathcal{R}_{a,b}$ denotes regions on the product of two rapidity tori in the obvious fashion. We will be most interested in the (bound state) dressing phase on the real line of the mirror theory, which lies in region $\mathcal{R}_{1,1}$. For completeness let us first briefly repeat the explicit proof that $\sigma$ satisfies the crossing equation, given in \cite{Hoare:2011wr}. In order to do so, we need to analytically continue the dressing phase to region $\mathcal{R}_{2,0}$.

\subsection{Proof of crossing}

\paragraph{The dressing phase in $\mathcal{R}_{1,0}$\\}
Continuing $\Phi$ through $|x_1^+|=1$ to $|x_1^+|<1$ gives

\begin{align}
{\cal R}_{1,0}:\quad \chi(x_1^+, x_2^\pm) =& \,\Phi(x_1^+, x_2^\pm) - \Psi(x_1^+, x_2^\pm)\,,\\
 \chi(x_1^-, x_2^\pm) =& \,\Phi(x_1^-, x_2^\pm)\,.
\end{align}

\paragraph{The dressing phase in $\mathcal{R}_{2,0}$\\}
With $|x_1^\pm|<1$ we have

\begin{align}
{\cal R}_{2,0}:\quad \chi(x_1^+, x_2^\pm) =& \,\Phi(x_1^+, x_2^\pm) - \Psi(x_1^+, x_2^\pm) +\frac{1}{i}\log \frac{\frac{1}{x_1^-}-x_2^\pm}{x_1^--x_2^\pm}\,,\\
\chi(x_1^-, x_2^\pm) =& \,\Phi(x_1^-, x_2^\pm) -
\Psi(x_1^-, x_2^\pm)\,.
\end{align}

\paragraph{Identities I\\}

In order to prove the crossing relation between $\mathcal{R}_{2,0}$ and $\mathcal{R}_{0,0}$ we will need some identities. First of all we have\footnote{These identities follow by changing variables from $z$ to $z^{-1}$ in the second integral on the left hand side.}
\begin{align}
\Phi(x_1,x_2) + \Phi(1/x_1,x_2) =&\, \Phi(0,x_2)\,,\label{eq:phibasic}\\
\Psi(x_1,x_2) + \Psi(x_1,1/x_2) =&\, \Psi(x_1,0)\,. \label{eq:psibasic}
\end{align}
Secondly we need
\begin{align}
\nonumber \Psi(1/x_1^-,x_2^+)-& \Psi(1/x_1^+, x_2^+) +\Psi(1/x_1^+, x_2^-)
-\Psi(1/x_1^-, x_2^-) =\\
& \frac{1}{i}\log\frac{1-\frac{1}{x_1^-x_2^+}}{1-\frac{1}{x_1^-x_2^-}}\frac{1-\frac{1}{x_1^+x_2^+}}{1-\frac{1}{x_1^+x_2^-}} + i \log \frac{1+\frac{\xi}{x_2^+}}{1+\frac{\xi}{x_2^-}}\, .
\nonumber
\end{align}
This identity follows from the identity
\begin{equation}
\label{eq:basicpsiid}
\Psi(1/x_1^-,x_2)-\Psi(1/x_1^+, x_2) = -i \log \frac{x_2 - \frac{1}{x_1^+}}{x_2}\frac{x_2 - \frac{1}{x_1^-}}{x_2+\xi} \, ,
\end{equation}
whose derivation plays an important role in the fusion of the mirror dressing phase below as well, so let us discuss it in some detail. To prove this identity, we begin by combining the two integrals
\begin{align}
\Psi(1/x_1^-,x_2)-& \Psi(1/x_1^+, x_2) =\\
& = i \oint_{|z|=1} \frac{dz}{2 \pi i} \frac{1}{z-x_2} \log  \frac{\Gamma_{q^2} (1+\frac{ig}{2}(u_1^- -u(z)))}{\Gamma_{q^2} (1-\frac{ig}{2}(u_1^--u(z)))}\frac{\Gamma_{q^2} (1-\frac{ig}{2}(u_1^+ -u(z)))}{\Gamma_{q^2} (1+\frac{ig}{2}(u_1^+-u(z)))}\nonumber
\end{align}
Next we use the defining property of the $\Gamma_q$ function given in eqn. \eqref{eq:Gammaqdef} to find
\begin{align}
\label{eq:basicpsiidstep2}
\Psi(1/x_1^-,x_2)-& \Psi(1/x_1^+, x_2) = \, i \oint_{|z|=1} \frac{dz}{2 \pi i} \frac{1}{z-x_2} \log \frac{1- q^{i g(u_1^- - u(z))}}{1-q^2}\frac{1- q^{i g(u(z) - u_1^+)}}{1-q^2}\, \nonumber\\
= & \, i \oint_{|z|=1} \frac{dz}{2 \pi i} \frac{1}{z-x_2} \log \left(1- \tfrac{z + \frac{1}{z} + \xi +\frac{1}{\xi}}{x_1^- + \frac{1}{x_1^-} + \xi +\frac{1}{\xi}}\right)\left(1- \tfrac{x_1^+ + \frac{1}{x_1^+} + \xi +\frac{1}{\xi}}{z + \frac{1}{z} + \xi +\frac{1}{\xi}}\right) \, , \\
= & \, i \oint_{|z|=1} \frac{dz}{2 \pi i} \frac{1}{z-x_2}
\log
\left(-\tfrac{x_1^-\xi}{(x_1^-+\xi)(x_1^-\xi+1)}
\tfrac{(z-x_1^-)\left(z-\tfrac{1}{x_1^-}\right)(z-x_1^+)\left(z-\tfrac{1}{x_1^+}\right)}{z (z+\xi)(z+\frac{1}{\xi})}
\right) \, ,\nonumber
\end{align}
where we note that the $\log(1-q^2)$ terms do not contribute to the integral as the integration contour can be shrunk to nothing. After integrating by parts\footnote{Note again that $|x_2|>1$.} we find
\begin{align}
\Psi(1/x_1^-,x_2)& \,-  \Psi(1/x_1^+, x_2) = \nonumber \\
& \, -i \oint_{|z|=1} \frac{dz}{2 \pi i} \log(z-x_2) \left( - \tfrac{1}{z} + \tfrac{1}{z-x_1^+} + \tfrac{1}{z-x_1^-} + \tfrac{1}{z-\frac{1}{x_1^+}} + \tfrac{1}{z-\frac{1}{x_1^-}} - \tfrac{1}{z+\xi} - \tfrac{1}{z+\frac{1}{\xi}}\right) \nonumber \\
& \, \hspace{67pt} = -i \log \frac{x_2 - \frac{1}{x_1^+}}{x_2}\frac{x_2 - \frac{1}{x_1^-}}{x_2+\xi} \, ,\label{eq:basicpsiidstep3}
\end{align}
by summing the poles within the unit circle. Here we used the fact that $|x_1^\pm|>1$. Had it been different, the respective $x$ would have to be replaced by $1/x$ in the above expression. Note that $\Psi$ is invariant under inversion of its first argument since $u(x) = u(1/x)$, so that the above also directly applies to $\Psi(x_1^-,x_2)-  \Psi(x_1^+, x_2)$.

In the construction of the mirror bound state dressing phase we will need a further identity\footnote{In these derivations we will not be very careful about factors of $i \pi$ - the typical ambiguity in defining $\tfrac{1}{2} \log(-1)^2$ - as they do not affect $\log \sigma^2$.}, namely
\begin{align}
\Psi(1/x_1^-,0) \,-  \Psi(1/x_1^+,0) =& \, i\log\frac{-\xi^2}{(1-q^2)^2} +i \log x_1^- x_1^+ \frac{x_1^-}{x_1^- + \xi}\frac{1}{\xi x_1^- + 1} \label{eq:basicpsiidx2zero}\\
 = & \, i\log\frac{-\xi^2}{(1-q^2)^2} +i \log x_1^- x_1^+  - i \log q^{-i g (u_1 -i/g)} (1-\xi^2)\, .\nonumber
\end{align}
which follows by carefully integrating \eqref{eq:basicpsiidstep2}, noting that now the constant log terms in the integral \emph{cannot} be dropped. In the limit $q\rightarrow1$, get the undeformed result
\begin{equation}
\Psi(1/x_1^-,0) \,-  \Psi(1/x_1^+,0) = i\log\frac{g^2}{4} +i \log x_1^- x_1^+
\end{equation}
since for $q\rightarrow1$, $\xi \rightarrow 0$ while $\tfrac{-\xi^2}{(1-q^2)^2}\rightarrow \tfrac{g^2}{4}$.

\paragraph{Crossing \\}

Putting the above together we can directly prove crossing between $\mathcal{R}_{0,0}$ and $\mathcal{R}_{2,0}$. This firstly makes use of the identity
\begin{equation}
\Phi(x_1,x_2) + \Phi(1/x_1,x_2) = \Phi(0,x_2)\,,
\end{equation}
valid for $|x_1| \neq 1$. Using this we find that
\begin{equation}
\Delta\tilde{\theta} \equiv \tilde{\theta}(z_1,z_2)+\tilde{\theta}(z_1+\omega_2,z_2)\, ,
\end{equation}
is given by
\begin{align}
\Delta\tilde{\theta} =& \,\Psi(1/x_1^-,x_2^+)-\Psi(1/x_1^+, x_2^+) +\Psi(1/x_1^+, x_2^-)
-\Psi(1/x_1^-, x_2^-)\nonumber  \\
&~~~~~~~~~~~~~~~~~+ \frac{1}{i}\log\frac{x_1^--x_2^+}{\frac{1}{x_1^-}-x_2^+}\frac{\frac{1}{x_1^-}-x_2^-}{x_1^--x_2^-}\,.
\end{align}
Then we use the identity
\begin{align}
\nonumber \Psi(1/x_1^-,x_2^+)-& \Psi(1/x_1^+, x_2^+) +\Psi(1/x_1^+, x_2^-)
-\Psi(1/x_1^-, x_2^-) =\\
& \frac{1}{i}\log\frac{1-\frac{1}{x_1^-x_2^+}}{1-\frac{1}{x_1^-x_2^-}}\frac{1-\frac{1}{x_1^+x_2^+}}{1-\frac{1}{x_1^+x_2^-}} + i \log \frac{1+\frac{\xi}{x_2^+}}{1+\frac{\xi}{x_2^-}}\, .
\nonumber
\end{align}
to finally find the crossing equation
\begin{align}
\nonumber \Delta\tilde{\theta} =& \, \frac{1}{i}\log\frac{1-\frac{1}{x_1^-x_2^+}}{1-\frac{1}{x_1^-x_2^-}}\frac{1-\frac{1}{x_1^+x_2^+}}{1-\frac{1}{x_1^+x_2^-}} +\frac{1}{i}\log\frac{x_1^--x_2^+}{\frac{1}{x_1^-}-x_2^+}\frac{\frac{1}{x_1^-}-x_2^-}{x_1^--x_2^-}+ i \log \frac{1+\frac{\xi}{x_2^+}}{1+\frac{\xi}{x_2^-}}\nonumber \\
= & \,
\frac{1}{i} \log \Big[\frac{x_2^- +\xi}{x_2^+ +\xi} h(x_1,x_2)\Big]\,.
\end{align}
In other words
\begin{equation}
\tilde{\sigma}(x_1,x_2) \tilde{\sigma}(1/x_1,x_2) = \frac{x_2^- +\xi}{x_2^+ +\xi} h(x_1,x_2)\, .
\end{equation}
If we now rewrite this in terms of $\sigma$ we get\footnote{Note that the dispersion implies $\frac{x_1^- +\xi}{x_1^+ +\xi}\frac{\frac{1}{x_1^-} +\xi}{\frac{1}{x_1^+} +\xi} = \frac{1}{q^2}$.}
\begin{equation}
\sigma(x_1,x_2)\sigma(1/x_1,x_2) = \frac{1}{q} \frac{x_2^-}{x_2^+} h(x_1,x_2)\, ,
\end{equation}
which is of course precisely the crossing equation $\sigma$ is supposed to solve.

\subsection{The improved bound state dressing phase of the mirror theory}

We would like to construct the dressing phase for bound states of the mirror theory. In order to do so we take the approach taken in \cite{Arutyunov:2009kf} in the undeformed case, and take the constituents of the mirror bound state such that only the first particle lies in region $\mathcal{R}_1$, i.e. $|x_1^-|>1$ but $|x_1^+|<1$, while all other particles lie in region $\mathcal{R}_0$ with $|x^\pm|>1$ \cite{BJ08}. In order to sum up all contributions to the bound state dressing phase of the mirror theory, in addition to the above we need the dressing phase in $\mathcal{R}_{1,1}$.

\paragraph{The dressing phase in $\mathcal{R}_{1,1}$\\}

Continuing the $\chi$ functions to the region with $|x_1^+|<1$, $|x_1^-|>1$, $|x_2^+|<1$ $|x_2^-|>1$ means we have to add additional terms to the expressions in region $\mathcal{R}_{1,0}$, corresponding to crossing $|x_2^+|=1$. This means the two $\Phi$ functions give extra $\Psi$ contributions, but also the already present $\Psi$ function gives an additional contribution. The full $\chi$ functions are then given by
\begin{align}
\nonumber {\cal R}_{1,1}:\quad  \chi(x_1^+, x_2^+) =&\,
\Phi(x_1^+, x_2^+)+ \Psi(x_2^+, x_1^+) - \Psi(x_1^+,
x_2^+)\\\nonumber &~~~~~~~~~~~~+i \log\frac{\Gamma_{q^2} (1+\frac{ig}{2}(x_1^++\frac{1}{x_1^+}-x_2^+-\frac{1}{x_2^+}))}
{\Gamma_{q^2} (1-\frac{ig}{2}(x_1^++\frac{1}{x_1^+}-x_2^+-\frac{1}{x_2^+}))}\,,\\\nonumber
\chi(x_1^+, x_2^-) =&\, \Phi(x_1^+, x_2^-) - \Psi(x_1^+, x_2^-)
\,,\\
\chi(x_1^-, x_2^+) =&\, \Phi(x_1^-, x_2^+)+ \Psi(x_2^+, x_1^-)\,,\nonumber\\
\chi(x_1^-, x_2^-) =&\, \Phi(x_1^-, x_2^-)\,,~~~~~~~~
\end{align}

\paragraph{The bound state dressing phase I\\}

As already indicated, the constituents of a mirror bound state can be taken to lie in the regions $\mathcal{R}_{1,1}$, $\mathcal{R}_{1,0}$, $\mathcal{R}_{0,1}$, and $\mathcal{R}_{0,0}$. By antisymmetry of the dressing factor the above sections provide the appropriate dressing factors for the constituents, which can then be fused. At this point we would like to introduce the improved dressing phase $\Sigma$
\begin{align}
\Sigma(x_1,x_2) \equiv & \, \sigma(x_1,x_2) \frac{1-\frac{1}{x_1^+x_2^-}}{1-\frac{1}{x_1^-x_2^+}} \\
=& \, e^{i\left(\chi(x_1^+,x_2^+) - \chi(x_1^-,x_2^+) - \chi(x_1^+,x_2^-) + \chi(x_1^-,x_2^-)\right)}  \sqrt{P(x_1)/P(x_2)} \frac{1-\frac{1}{x_1^+x_2^-}}{1-\frac{1}{x_1^-x_2^+}} \, , \nonumber
\end{align}
where $P$ is defined in \eqref{eq:dressingvsdressing}. The chosen representative constituents for a $Q$-particle bound are parametrized as
\begin{align}
x_j^-(u) = &\, x_s(u+(Q-2j)\frac{i}{g}) \, , \, \, \, j= 1,\ldots,Q\, , \\
x_1^+(u) = &\, \frac{1}{x_s(u+ i Q/g)}\, , \, \, \, x_j^+(u) = x_s(u+(Q-2j+2)\frac{i}{g})\, .
\end{align}
We then denote the final mirror theory bound state parameters as
\begin{equation}
y_1^\pm = x_{1/Q}^\pm(u) \, , \, \, \, y_2^\pm = x_{1/Q^\prime}^\pm(v) \, .
\end{equation}

Summing up all contributions for the $Q$-particle bound state - $Q^\prime$-particle bound state improved mirror dressing phase using the appropriate form of the $\chi$ functions in each region, we directly find
\begin{align}
-i\log\Sigma^{QQ'}(y_1,y_2) =& \,
\Phi(y_1^+,y_2^+)-\Phi(y_1^+,y_2^-)-\Phi(y_1^-,y_2^+)+\Phi(y_1^-,y_2^-)
\nonumber\\ &\,-
\Psi(y_1^+,y_2^+)+\Psi(y_1^+,y_2^-)+\Psi(y_{2}^+,y_1^+)-\Psi(y_{2}^+,y_1^-)
\nonumber\\ &\,-i\log \frac{\Gamma_{q^2}\left(1-\frac{ig}{2}\left(y_1^++\frac{1}{y_1^+}-y_2^+-\frac{1}{y_2^+}\right)\right)}
{\Gamma_{q^2}\left(1+\frac{ig}{2}\left(y_1^++\frac{1}{y_1^+}-y_2^+-\frac{1}{y_2^+}\right)\right)} P(y_1)^{Q^\prime/2}P(y_2)^{-Q/2}
\nonumber\\
&\,-i\log\frac{1- \frac{1}{y_1^+y_2^-}}{1-\frac{1}{y_1^-y_2^+}}\prod_{j=1}^{Q-1} \frac{ 1- \frac{1}{x_j^-y_2^-}}{1-\frac{1}{x_j^-y_2^+}} \prod_{k=1}^{Q'-1}\frac{ 1- \frac{1}{y_1^+z_k^-}}{ 1-\frac{1}{y_1^-z_k^-}}\,,\label{eq:improveddressingphasedirect}
\end{align}
where we emphasize that
\begin{equation}
|y_1^+| <1\, , \, \, \, |y_1^-| >1\, , \, \, \, |y_2^+| <1\, , \, \, \, |y_2^-| >1\,.
\end{equation}
This result has an apparent dependence on the bound state constituents. To manifestly remove this dependence we need a few more identities.

\paragraph{Identities II\\}

Firstly we have
\begin{align}
\Psi(y_{1}^+, y_2^-) -\Psi(y_{1}^-, y_2^-) =& \, i\log\frac{y_2^- - y_1^+}{y_2^-} \frac{y_2^- -\frac{1}{y_1^-}}{y_2^- + \xi} \prod_{j=1}^{Q-1}\frac{y_2^--\frac{1}{x_j^-}}{y_2^-}\frac{y_2^--\frac{1}{x_j^-}}{y_2^- + \xi} \label{eq:psiboundprop}\\
= & \,i\log \left(\frac{y_2^-}{y_2^- + \xi}\right)^Q \left(1-\frac{y_1^+}{y_2^\pm} \right)\left(1-\frac{1}{y_1^- y_2^-}\right)\prod_{j=1}^{Q-1}\left(1-\frac{1}{x_j^-y_2^-}\right)^2 \, , \nonumber
\end{align}
which follows by applying the defining property of $\Gamma_q$ $Q$ times in the derivation (\ref{eq:basicpsiid},\ref{eq:basicpsiidstep2},\ref{eq:basicpsiidstep3}) and using the fact that $|y_1^+|<1$ while $|y_{1,2}^-|>1$. With both $y_1$ and $y_2$ referring to particles in region $\mathcal{R}_1$ this also immediately implies
\begin{equation}
\Psi(y_2^+,y_1^-) -\Psi(y_2^-,y_1^-) =i\log \left(\frac{y_1^-}{y_1^- + \xi}\right)^{Q^\prime} \left(1-\frac{y_2^+}{y_1^-} \right)\left(1-\frac{1}{y_1^- y_2^-}\right)\prod_{j=1}^{Q^\prime-1}\left(1-\frac{1}{x_j^-y_1^-}\right)^2 \, .\nonumber
\end{equation}
Then we also have the bound-state analogue of \eqref{eq:basicpsiidx2zero} for a particle in region $\mathcal{R}_1$
\begin{align}
\Psi(y_1^-,0) \,-  \Psi(y_1^+,0) =& \, i Q \log\frac{-\xi^2}{(1-q^2)^2} +i \log \frac{y_1^-}{y_1^+}
 \prod_{j=1}^{Q-1} (x_j^-)^2 - i Q \log q^{-i g(u-\tfrac{i}{g})} (1-\xi^2)\, .\nonumber
\end{align}
By applying the basic property \eqref{eq:psibasic} we can reexpress terms of the type $\Psi(y_1^+,y_2^+)$ and apply the above type of identities to find
\begin{align}
\Psi(y_1^+,y_2^+) - \Psi(y_1^-,y_2^+) = & \, \Psi(y_1^-,\tfrac{1}{y_2^+}) - \Psi(y_1^+,\tfrac{1}{y_2^+}) + \Psi(y_1^+,0) - \Psi(y_1^-,0)\nonumber \\
= & \, i Q \log (\xi y_2^+ +1) (1-q^2)^2(1-\xi^{-2}) q^{-i g(u-\tfrac{i}{g})} + \nonumber \\
 & \, - i\log (y_2^+ -y_1^-)\left(y_2^+-\frac{1}{y_1^+}\right)\prod_{j=1}^{Q-1}(x_j^--y_2^+)^2 \nonumber\\
= &\, i Q \log (\xi y_2^+ +1) (1-q^2)^2(1-\xi^{-2}) q^{-i g(u-\tfrac{i}{g})}+ \nonumber \\
 & \, - i\log (y_2^+ -y_1^-)\left(y_2^+-\frac{1}{y_1^+}\right)\prod_{j=1}^{Q-1}\left(1-\frac{1}{x_j^-y_2^+}\right)^{-2} \\
 &\, - i \log \prod_{j=1}^{Q-1}(\xi^{-1} - \xi)^2 \left(q^{-i g(u+(Q-2j)\tfrac{i}{g})}-q^{-ig(v+Q^\prime\tfrac{i}{g})}\right)^2\, .\nonumber
\end{align}
Similarly we find\footnote{Here the parameters $x_j^-$ of course refer to bound state number $Q^\prime$ and center $v$.}
\begin{align}
\Psi(y_2^+,y_1^+) - \Psi(y_2^-,y_1^+)
= &\, i Q^\prime \log (\xi y_1^+ +1) (1-q^2)^2(1-\xi^{-2}) q^{-i g(v-\tfrac{i}{g})} + \nonumber \\
 & \, - i\log (y_1^+ -y_2^-)\left(y_1^+-\frac{1}{y_2^+}\right)\prod_{j=1}^{Q^\prime-1}\left(1-\frac{1}{x_j^-y_1^+}\right)^{-2} \\
 &\, - i \log \prod_{j=1}^{Q^\prime -1}(\xi^{-1} - \xi)^2 \left(q^{-ig(v+(Q^\prime-2j)\tfrac{i}{g})}-q^{-i g(u+Q\tfrac{i}{g})}\right)^2\nonumber
\, .
\end{align}
Putting the above identities together with \eqref{eq:psiboundprop} we have
\begin{align}
-\Psi(y_1^+,y_2^+)& \, +\Psi(y_1^+,y_2^-)+\Psi(y_{2}^+,y_1^+)-\Psi(y_{2}^+,y_1^-)
-i\log \prod_{j=1}^{Q-1} \frac{1- \frac{1}{x_j^-y_2^-}}{1-\frac{1}{x_j^-y_2^+}} \prod_{k=1}^{Q'-1}\frac{ 1- \frac{1}{y_1^+z_k^-}}{ 1-\frac{1}{y_1^-z_k^-}} \nonumber \\
 = & \,\frac{1}{2}\left(-\Psi(y_1^+,y_2^+)\, +\Psi(y_1^+,y_2^-)+\Psi(y_{2}^+,y_1^+)-\Psi(y_{2}^+,y_1^-)\right) \nonumber \\
 & \, + \frac{1}{2}\left(-\Psi(y_1^-,y_2^+)\, +\Psi(y_1^-,y_2^-)+\Psi(y_{2}^-,y_1^+)-\Psi(y_{2}^-,y_1^-)\right)\nonumber \\
 & \, + i \log \frac{\prod_{j=1}^{Q-1} \left(q^{-i g(u+(Q-2j)\tfrac{i}{g})}-q^{-ig(v+Q^\prime\tfrac{i}{g})}\right)/(1-q^2)}{\prod_{j=1}^{Q^\prime-1} \left(q^{-ig(v+(Q^\prime-2j)\tfrac{i}{g})}-q^{-i g(u+Q\tfrac{i}{g})}\right)/(1-q^2)} -\frac{i}{2} \log \frac{y_1^+}{y_1^-} \frac{y_2^-}{y_2^+}\nonumber \\
 & \, + \frac{i Q^\prime}{2} \log q^{-ig(v-\tfrac{i}{g})} (\xi y_1^+ + 1) \frac{y_1^- + \xi}{y_1^-} - \frac{i Q}{2} \log q^{-ig(u-\tfrac{i}{g})} (\xi y_2^+ + 1) \frac{y_2^- + \xi}{y_2^-} \nonumber \\
 & \, + i \frac{Q-Q^\prime}{2} \log(\xi^2-1)\, .
\end{align}
Then we can use the defining property of the $\Gamma_q$ function to find
\begin{align}
& \frac{\prod_{j=1}^{Q^\prime-1} \left(q^{-ig(v+(Q^\prime-2j)\tfrac{i}{g})}-q^{-i g(u+Q\tfrac{i}{g})}\right)/(1-q^2)}{\prod_{j=1}^{Q-1}   \left(q^{-i g(u+(Q-2j)\tfrac{i}{g})}-q^{-ig(v+Q^\prime\tfrac{i}{g})}\right)/(1-q^2)}\frac{\Gamma_{q^2}\left(1- \frac{ig}{2}(u(y_1^+) - u(y_2^+))\right)}{\Gamma_{q^2}\left(1+ \frac{ig}{2}(u(y_1^+) - u(y_2^+))\right)}\nonumber\\
& \hspace{50pt} = \frac{\prod_{j=1}^{Q^\prime-1} q^{-ig(v+(Q^\prime-2j)\tfrac{i}{g})}}{\prod_{j=1}^{Q-1}   q^{-i g(u+(Q-2j)\tfrac{i}{g})}} \frac{\Gamma_{q^2}\left(Q^\prime - \frac{ig}{2}(u(y_1^+) - u(y_2^+))\right)}{\Gamma_{q^2}\left(Q+ \frac{ig}{2}(u(y_1^+) - u(y_2^+))\right)} \nonumber\\
& \hspace{50pt} = q^{-ig (Q^\prime-1) v} q^{ig (Q-1) u} \frac{\Gamma_{q^2}\left(Q^\prime - \frac{ig}{2}(u(y_1^+) - u(y_2^+))\right)}{\Gamma_{q^2}\left(Q+ \frac{ig}{2}(u(y_1^+) - u(y_2^+))\right)} \, .
\end{align}

\paragraph{The bound state dressing phase II\\}

By using the above identities, the improved dressing phase given in eq. \eqref{eq:improveddressingphasedirect} can be written as
\begin{align}
-i\log\Sigma^{QQ'}(y_1,y_2) =& \, \Phi(y_1^+,y_2^+)-\Phi(y_1^+,y_2^-)-\Phi(y_1^-,y_2^+)+\Phi(y_1^-,y_2^-) \label{eq:improveddressingphase} \\ &\, -\frac{1}{2}\left(\Psi(y_1^+,y_2^+)+\Psi(y_1^-,y_2^+)-\Psi(y_1^+,y_2^-)-\Psi(y_1^-,y_2^-)\right) \nonumber \\
&\, +\frac{1}{2}\left(\Psi(y_{2}^+,y_1^+)+\Psi(y_{2}^-,y_1^+)-\Psi(y_{2}^+,y_1^-)
-\Psi(y_{2}^-,y_1^-) \right) \nonumber \\
&\, - i \log \frac{i^Q \, \Gamma_{q^2}\left(Q^\prime - \frac{ig}{2}(u(y_1^+) - u(y_2^+))\right)}{i^{Q^\prime} \Gamma_{q^2}\left(Q+ \frac{ig}{2}(u(y_1^+) - u(y_2^+))\right)}\frac{1- \frac{1}{y_1^+y_2^-}}{1-\frac{1}{y_1^-y_2^+}} \sqrt{\frac{y_1^+}{y_1^-} \frac{y_2^-}{y_2^+}} \nonumber \\
& \, + \frac{i}{2} \log q^{Q-Q^\prime } q^{-i g (Q+Q^\prime -2)(u-v)}\nonumber \, .
\end{align}
This version of the improved mirror bound state dressing phase manifests its proper fusion, as any apparent dependence on the bound state constituents has been removed.

To see that the improved dressing phase is unitary it is important to realize that the $\log \Gamma_{q^2}$ combinations entering in $\Phi$, $\Psi$ and explicitly in the formula above, are \emph{not} real. Rather their imaginary parts cancel precisely against the manifestly imaginary terms in the above prescription. One way to show this is to realize that the above expression is independent of the constituent particles, and so should agree with the improved bound state dressing phase obtained by fusing over a manifestly mirror theory conjugation symmetric bound state configuration. The latter expression can be easily proven to be unitary by using the crossing equation and the generalized unitarity of the dressing phase about the string line. We have also verified unitarity of the above expression explicitly numerically. The further relevant analytic properties of the dressing phase are covered in the section below. Note that this expression for the dressing phase manifestly reduces to the improved mirror dressing phase of the undeformed theory.

\subsection{Action of the discrete Laplace operator}

The mirror-mirror bound state dressing phase is a holomorphic function in the intersection of the region $R_{1,1}$ and the mirror region, which in particular contains the mirror line. This follows by the discontinuity relations for the $\Psi$ functions in their first argument, which we discuss below. Holomophicity and fusion mean that the dressing phase is immediately annihilated by the discrete Laplace operator $\Delta_{Q^\prime,P} = \delta_{Q^\prime,P}s^{-1} -( \delta_{Q^\prime +1,P}+\delta_{Q^\prime-1,P})$ for $P\geq 1$, since we are far from any branch cut ambiguities of the arguments of the dressing phase. For $P=1$ this is no longer the case; acting with the operator
\begin{equation}
\label{eq:Kplus1invQ1}
\Delta_{Q^\prime,1}(v) = \delta_{Q^\prime,1}s^{-1}(v) - \delta_{Q^\prime,2}\, ,
\end{equation}
only annihilates the mirror-mirror bound state dressing phase for $v \in (-u_b,u_b)$\footnote{The significance of the interval $(-u_b,u_b)$ lies in the fact that $x_m(u)$ covers the unit circle precisely as $u$ runs over this interval.}. This is the reason the $Y$-system for $Y_1$ only exists on this interval of the real mirror line. Because we do not encounter any branch cuts of the arguments of the dressing phase for $v \in (-u_b,u_b)$, annihilation within this interval still follows immediately by holomorphicity.

To show that the dressing phase is holomorphic in the intersection of $\mathcal{R}_{1,1}$ and the mirror region, in addition to the formulae already derived we will need the discontinuity of the $\Psi$ function $\Psi(x_1,x_2)$ in its first argument. The relevant discontinuity is at $u_1 = u + 2 i /g$ with $u\in (-u_b,u_b)$. To find it we proceed as in the undeformed case \cite{Arutyunov:2009kf} and integrate by parts to write the $\Psi$ function as
\begin{equation}
\Psi(x_1,x_2) = -\frac{g}{2} \oint_{|z|=1} \frac{dz}{2 \pi i} \log(z-x_2)\frac{d u}{dz}\left( \psi_{q^2}(1+\tfrac{i g }{2}(u_1-u(z)))+\psi_{q^2}(1-\tfrac{i g }{2}(u_1-u(z)))\right)\, ,\nonumber
\end{equation}
where $\psi_q$ is the $q$-digamma function, the logarithmic derivative of the $\Gamma_q$ function. From the defining relation of the $\Gamma_q$ function it is easy to see that the $q$-digamma function still has simple poles at the negative integers with residue negative one. This makes the discontinuities immediately clear as two poles\footnote{Recall that $u(z) = u(1/z)$.} hit the integration contour when $u_1 = u + 2 i /g$. The corresponding discontinuity is then simply
\begin{equation}
\label{eq:Psidiscfirstarg}
\lim_{\epsilon \rightarrow 0} \Psi(e^\epsilon x_1,x_2) - \Psi(e^{-\epsilon} x_1,x_2) = - i \log \frac{x_m(u) - x_2}{\frac{1}{x_m(u)} -x_2} \, ,
\end{equation}
following from the two pole contributions just inside and just outside the unit circle\footnote{Effectively giving $x_m$ and $1/x_m$ respectively.} respectively. Analogous formulae apply for discontinuities at $u_1 = u + 2 n i/g$, exactly as in the undeformed case \cite{Arutyunov:2009kf}. For example for the discontinuity through $u_1 = u - 2 i/g$ we get
\begin{equation}
\lim_{\epsilon \rightarrow 0} \Psi(e^\epsilon x_1,x_2) - \Psi(e^{-\epsilon} x_1,x_2) =  i \log \frac{x_m(u) - x_2}{\frac{1}{x_m(u)} -x_2} \, ,
\end{equation}
where the different sign arises because the poles cross the integration contour in the opposite direction. This shows that the cuts of combinations like $\Psi(y_2^+,y_1^+)+\Psi(y_2^-,y_1^+)$ precisely cancel in the intersection of $R_{1,1}$ and the mirror region, making the dressing phase a holomorphic function there as mentioned above.

With this identity we can also directly compute the action of $\Delta_{Q^\prime,1}(v)$ on $\Sigma^{QQ^\prime}(u,v)$ almost identically to the undeformed case \cite{AF09d}, and show explicitly that it indeed vanishes when $v \in (-u_b,u_b)$. Acting on the $\Phi$ functions, noting that the $\Psi$ function is precisely its discontinuity when its second argument crosses the unit circle (cf. eq. \eqref{eq:phidef}), we get
\begin{equation}
\left(\Phi(y_1^+,y_2^+) -\Phi(y_1^+,y_2^-)-\Phi(y_1^-,y_2^+)+\Phi(y_1^-,y_2^-)\right)\Delta_{Q^\prime,1}(v) = \Psi(x_m(v),y_1^-)-\Psi(x_m(v),y_1^+)\, .
\end{equation}
For the first line of the $\Psi$ functions in \eqref{eq:improveddressingphase} we get
\begin{align}
-\frac{1}{2}\left(\Psi(y_1^+,y_2^+) +\Psi(y_1^-,y_2^+)-\Psi(y_1^+,y_2^-)-\Psi(y_1^-,y_2^-)\right)\Delta_{Q^\prime,1}(v)& =  \\
= -\frac{i}{2}\log q^{-ig(Q-2)(u-v)} S_Q + i \log i^Q & \frac{\Gamma_q^2\left(\tfrac{Q}{2}-\tfrac{i g}{2}(u-v)\right)}{\Gamma_q^2\left(\tfrac{Q}{2}+\tfrac{i g}{2}(u-v)\right)}\nonumber
\end{align}
which simply follows from the integral representation for $\Psi$ of eq. \eqref{eq:psidef} when its second argument crosses the unit circle. $\Delta$ effectively acts on the first argument of the second line of $\Psi$ functions, giving
\begin{align}
\frac{1}{2}\left(\Psi(y_2^+,y_1^+) +\Psi(y_2^-,y_1^+)-\Psi(y_2^+,y_1^-)-\Psi(y_2^-,y_1^-)\right)\Delta_{Q^\prime,1}(v) & = \nonumber\\
= \Psi(x_m(v),y_1^+)-\Psi(x_m(v),y_1^-) - \frac{i}{2} \frac{y_1^+ - x_m(v)}{y_1^+ - \frac{1}{x_m(v)}}&\frac{y_1^- - \frac{1}{x_m(v)}}{y_1^- - x_m(v)}\, .
\end{align}
where we used the discontinuity of the $\Psi$ function we just computed, given by eq. \eqref{eq:Psidiscfirstarg}. This can be partly rewritten as
\begin{equation}
-\frac{i}{2} \log \frac{y_1^+ - x_m(v)}{y_1^+ - \frac{1}{x_m(v)}}\frac{y_1^- - \frac{1}{x_m(v)}}{y_1^- - x_m(v)} = i
\log \frac{y_1^+ - \frac{1}{x_m(v)}}{y_1^- - \frac{1}{x_m(v)}}\sqrt{\frac{y_1^-}{y_1^+}} +\frac{i}{2} \log q^{-Q}S_Q(u,v)\, ,
\end{equation}
which follows from the general identity
\begin{equation}
\frac{1 - \tfrac{1}{x^-_i x^-_j}}{1- \frac{1}{x^+_i x^+_j}}\frac{x^-_i - x^-_j}{x^+_i - x^+_j} = q^{-(Q+M)} S_{Q-M} \, ,
\end{equation}
where particles $i$ and $j$ have bound state numbers $Q$ and $M$ respectively. Finally the last terms give
\begin{align}
-i \log \frac{i^Q \, \Gamma_{q^2}\left(Q^\prime - \frac{ig}{2}(u(y_1^+) - u(y_2^+))\right)}{i^{Q^\prime} \Gamma_{q^2}\left(Q+ \frac{ig}{2}(u(y_1^+) - u(y_2^+))\right)}\frac{1- \frac{1}{y_1^+y_2^-}}{1-\frac{1}{y_1^-y_2^+}} \sqrt{\frac{y_1^+}{y_1^-} \frac{y_2^-}{y_2^+}} \Delta_{Q^\prime,1}(v) &=\nonumber\\
= -i \log i^Q \frac{\Gamma_q^2\left(\tfrac{Q}{2}-\tfrac{i g}{2}(u-v)\right)}{\Gamma_q^2\left(\tfrac{Q}{2}+\tfrac{i g}{2}(u-v)\right)} - i \log \frac{y_1^+ - \frac{1}{x_m(v)}}{y_1^- - \frac{1}{x_m(v)}} & \sqrt{\frac{y_1^-}{y_1^+}}
\end{align}
and
\begin{equation}
\frac{i}{2} \log q^{Q-Q^\prime } q^{-i g (Q+Q^\prime -2)(u-v)}\Delta_{Q^\prime,1}(v) = \frac{i}{2} \log q^Q  q^{-i g (Q-2)(u-v)}
\end{equation}
respectively. Adding everything up we get zero as promised. In summary, this means
\begin{equation}
\Sigma^{Q Q^\prime} \Delta_{Q^\prime,1}(u,v) = 0 \, , \, \, \, \, \, \, \, \mbox{for} \, \, v \in (-u_b,u_b)\,.
\end{equation}
When $v$ is outside this interval the result of applying $\Delta$ is nonzero. In the undeformed case the resulting kernel can be cast in a simple form \cite{AF09d}, we will not pursue a simple version of the resulting deformed kernel here.

\end{document}